\documentclass[useAMS,usenatbib]{mnras}
\usepackage{amsmath}
\usepackage{amssymb}
\usepackage{graphicx}
\usepackage{xcolor}
\usepackage{hhline}
\usepackage{cancel}
\usepackage[normalem]{ulem}

\newcommand{\be}{\begin{equation}}
\newcommand{\ee}{\end{equation}}
\newcommand{\Pn}{\mathcal{P}(N)}
\newcommand{\p}{\mathcal{P}}
\newcommand{\e}{\times10^}

\newcommand{\pa}{\partial}
\newcommand{\Nbar}{\overline{N}}
\newcommand{\pmag}{\p_{\rm{mag}}}
\newcommand{\thb}{\mathbf{\theta}}

\title[PDFs of intensity maps]{Insights from probability distribution functions of intensity maps}
\author[Patrick C. Breysse et al.]{Patrick C. Breysse,$^{1}$\thanks{pbreysse@pha.jhu.edu (PCB)}
Ely D. Kovetz,$^{1}$, 
Peter S. Behroozi,$^{2,3}$\, 
Liang Dai$^{4,5}$, 
\newauthor and Marc Kamionkowski$^{1}$\footnotemark[1] \\
$^{1}$ Department of Physics and Astronomy, Johns Hopkins University, Baltimore, MD 21218 USA\\
$^{2}$ Hubble Fellow\\
$^{3}$ Astronomy Department, University of California, Berkeley, Berkeley CA 94720, USA\\
$^{4}$ Einstein Fellow\\
$^{5}$ Institute for Advanced Study, Einstein Drive, Princeton, New Jersey 08540, USA}

\begin{document}
\label{firstpage}
\pagerange{\pageref{firstpage}--\pageref{lastpage}}
\maketitle

\begin{abstract}
In the next few years, intensity-mapping surveys that target lines such as CO, Ly$\alpha$, and CII stand to provide powerful probes of high-redshift astrophysics. However, these line emissions are highly non-Gaussian, and so the typical power-spectrum methods used to study these maps will leave out a significant amount of information.  We propose a new statistic, the probability distribution of voxel intensities, which can access this extra information.  Using a model of a CO intensity map at $z\sim3$ as an example, we demonstrate that this voxel intensity distribution (VID) provides substantial constraining power beyond what is obtainable from the power spectrum alone.  We find that a future survey similar to the planned COMAP Full experiment could constrain the CO luminosity function to order $\sim10\%$.  We also explore the effects of contamination from continuum emission, interloper lines, and gravitational lensing on our constraints and find that the VID statistic retains significant constraining power even in pessimistic scenarios.
\end{abstract}


\section{Introduction}
Intensity mapping has arisen in recent years as a powerful new means to observe the high-redshift universe.  Traditional galaxy surveys become less effective at great distances as more luminous galaxies start to fall below detection thresholds.  intensity-mapping surveys, first proposed by \citet{Suginohara1999}, make use of the emission from these fainter galaxies by measuring the intensity fluctuations of a chosen spectral line on large spatial scales.  Such a survey thus makes use of the aggregate emission from all of the galaxies within a target volume, and can make statistical measurements of the entire galaxy population.  By studying a single emission line, it is possible to observe these intensity fluctuations in three dimensions, as the observed frequency of a line maps one-to-one to the emission redshift.  This allows detailed study of line emission in relatively unexplored periods of cosmic history.

The fluctuations observed in an intensity mapping survey depend on the luminosity function and spatial distribution of the source galaxies.  The luminosity function depends on the detailed astrophysical conditions within the emitters, such as star formation rates and metallicities, while the spatial distribution traces the underlying dark matter field, the properties of which in turn depend on cosmological parameters.  Intensity mapping can thus provide information about a wide variety of cosmological and astrophysical topics.  Typically, all of this information is extracted from a map using its power spectrum, a powerful statistic which has proven valuable for studying both galaxy distrubutions \citep{Tegmark1998} and the cosmic microwave background \citep{Planck2015c}.

However, for highly non-Gaussian fields, the power spectrum, a two-point statistic, leaves out a significant amount of information.  In the CMB, non-Gaussianities are typically probed using higher-point statistics such as the bispectrum and trispectrum \citep{Bartolo2010}.  Unfortunately, these statistics are rather difficult to work with, both from a theoretical and observational perspective.  We propose instead to study the one-point statistics of intensity maps using a quantity we will refer to as the voxel intensity distribution, or VID.  This VID statistic is the probability distribution function of observed pixel intensities.  It can be predicted in a straightforward manner from a model luminosity function, and it can be estimated from a map simply by making a histogram of the observed intensity values.  

Our VID method is an extension of a technique known as probability of deflection, or $P(D)$ analysis, which is a general method for predicting observed intensities from confusion-limited populations. $P(D)$ analysis was originally developed for radio astronomy \citep{Scheuer1957}, but has since been applied to observations ranging from gamma rays \citep{Lee2009,Lee2015} to X-rays \citep{Barcons1994} to the infrared \citep{Glenn2010}.  Since intensity maps provide deliberately confused observations of galaxy populations, they are good candidates for $P(D)$ analysis.  \citet{Breysse2016a} first discussed this technique in the context of intensity mapping as a method to measure high-redshift star formation rates.  Here, we study this method in far more detail with the goal of creating a procedure that can be readily applied to many different intensity-mapping surveys targeting different lines.

The most well-known line used for intensity mapping is the 21 cm spin-flip line from neutral hydrogen (see for example \citet{Morales2010} and references therein).  Experiments such as the Precision Array for Probing the Epoch of Reionization (PAPER, \citealt{Ali2015}), the Murchison Widefield Array (MWA, \citealt{Tingay2013}), the LOw-Frequency ARray (LOFAR, \citealt{Haarlem2013}), the Hydrogen Epoch of Reionization Array (HERA, \citealt{DeBoer2016}) and the Square Kilometer Array (SKA, \citealt{Santos2015}) seeking to study the epoch of reionization and experiments like the Canadian Hydrogen Intensity Mapping Experiment (CHIME, \citealt{Bandura2014}) and the Hydrogen Intensity and Real-time Analysis eXperiment (HIRAX, \citealt{Newburgh2016}) seeking to study galaxies around $z\sim1$.  However, different lines probe different astrophysical processes, and have to deal with vastly different foregrounds and systematic effects, so there has been a recent effort to study lines besides 21 cm.

The Lyman $\alpha$ line \citep{Pullen2014,Comaschi2016}, targeted by the Spectro-Photometer for the History of the Universe, Epoch of Reionization, and Ices Explorer (SPHEREx, \citealt{Dore2014}), also traces hydrogen gas, but in hotter environments than 21 cm.  Ionized regions can be studied using lines such as the 158 $\mu$m CII fine structure line \citep{Yue2015,Silva2015}, using experiments such as the planned Tomographic Intensity Mapping Experiment (TIME, \citealt{Crites2014}).  Rotational transitions of CO molecules, sought by the CO Power Spectrum Survey (COPSS, \citep{Keating2016}) and CO Mapping Array Pathfinder (COMAP, \citep{Li2016}), probe cool, dense molecular gas \citep{Righi2008,Pullen2013,Breysse2014}.  Many other lines, including Balmer series and molecular hydrogen lines as well as lines from helium, oxygen, and nitrogen have also been discussed in the literature \citep{Visbal2010,Gong2013,Visbal2015,Fonseca2016}.

Below, we provide a detailed formalism for the VID statistic that should be readily applicable to a wide variety of intensity mapping models.  In order to demonstrate the efficacy of this method, we apply this formalism to a four-parameter model of a CO intensity mapping survey.  We apply a Fisher matrix analysis to this model and demonstrate that the VID can constrain the parameters of this model to order $\sim10\%$.  We then go on to consider several forms of foreground contamination that are expected to affect intensity maps and find that the VID retains its usefulness even under rather pessimistic assumptions.

For this work we assume a $\Lambda$CDM cosmology with $(\Omega_m,\Omega_\Lambda,h,\sigma_8,n_s)=[0.27,0.73,0.7,0.8,0.96]$ which is consistent with the WMAP results.  Section 2 contains a discussion of the power spectrum and its limitations along with the presentation of our VID formalism.  Section 3 describes our CO emission model, which we use in Section 4 to demonstrate the constraining power of the VID.  Section 5 investigates how contamination from continuum emission, interloper lines, and gravitational lensing effects our constraints.  We discuss our results in detail in Section 6 and conclude in Section 7.

\section{Formalism}
Here we will discuss the method we use to apply $P(D)$ analysis techniques to intensity maps.  We will begin with a brief discussion of the well-known formalism for the power spectrum of such a map, and demonstrate its limitations when attempting to constrain luminosity functions.  Then, we will derive our proposed VID statistic in detail and illustrate why it contains information beyond that found in the power spectrum.  Our formalism in this section is entirely independent of which line or luminosity function model is used, so it is readily applicable to many different intensity mapping experiments.

\subsection{Power spectrum}
The primary statistic used to date when discussing intensity maps is the power spectrum $P(k)$.  The power spectrum is the Fourier transform of the two-point correlation function, and is thus a two-point statistic.  An intensity map of a line emitted at redshift $z$ from galaxies with luminosity function $\Phi(L)$ has a power spectrum
\be
P(k,z) = \overline{T}^2(z)\overline{b}^2(z)P_m(k,z)+P_{\rm{shot}}(z),
\label{Pk}
\ee
(see, for example, \citet{Lidz2011,Breysse2014}, \citet{Gong2012}) The source galaxies trace the underlying dark matter distribution, which has a linear power spectrum $P_m(k,z)$.  The luminosity-weighted bias $\overline{b}$ of the galaxy population is given by
\be
\overline{b}=\frac{\int L b(L)\Phi(L)dL}{\int L\Phi(L)},
\label{bmean}
\ee
where $b(L)$ is the bias of a galaxy with luminosity $L$.  Many intensity mapping models assume a relation $L(M)$ between halo mass and line luminosity.  In these situations, one can replace Equation (\ref{bmean}) with mass integrals over some mass function $dn/dM$ and a mass-dependent bias $b(M)$.

The galaxy power spectrum $P_{\rm{gal}}(k)=\overline{b}^2P_m(k)$ is weighted by the square of the sky-averaged intensity $\overline{T}$, here written as a brightness temperature, to produce the intensity power spectrum.  The average intensity is related to the luminosity function by
\be
\begin{split}
\overline{T}&=\frac{c^3(1+z)^2}{8\pi k_B \nu_{\rm{em}}^3 H(z)}\int_0^\infty L \Phi(L)dL \\
& \equiv X_{LT}(z)\int_0^\infty L \Phi(L)dL,
\end{split}
\label{Tavg}
\ee
where $c$ is the speed of light, $k_B$ is Boltzmann's constant, $H(z)$ is the Hubble parameter at redshift $z$, and we have defined $X_{LT}(z)$ to keep our notation compact.  Because the emission in an intensity map comes from a population of discrete sources, there is an additional scale-independent shot noise component $P_{\rm{shot}}$ in the power spectrum that is given by
\be
P_{\rm{shot}}=X_{LT}^2\int_0^\infty L^2\Phi(L)dL.
\label{Pshot}
\ee

The power spectrum is a useful statistic when studying cosmological density fields, but it suffers from one key limitation.  All of the information about a random field is contained within its power spectrum if and only if the field is perfectly Gaussian.  However, we expect the small-scale fluctuations in an intensity map to be highly non-Gaussian, as the measured intensity is the product of highly nonlinear processes within the galaxy population.  Thus, the power spectrum alone misses out on much of the information content of a map.  

This can be easily seen by looking at Equations (\ref{Tavg}) and (\ref{Pshot}).  The power spectrum depends only on the first two moments of the luminosity function $\Phi(L)$. No higher moments can be measured from this statistic.  As with CMB measurements, higher moments may be measurable using higher order $n$-point statistics such as the bispectrum, but these are computationally difficult \citep{Planck2015}.  One could also obtain additional information through cross-correlations of different lines, but while this allows the study of different properties of the galaxy distribution \citep{Breysse2016b}, it adds little information about the initial target line.  A much more straightforward method would be to consider instead the one-point statistics of an intensity map, as described in \citet{Breysse2016a}.

\subsection{Voxel Intensity Distribution}
We derive the VID based on the $P(D)$ computation presented in \citet{Lee2009}, modified somewhat to include the effects of clustering.  The formalism presented here is an expanded version of the one described in \citet{Breysse2016a}.  

Consider a volume of space at redshift $z$ containing a population of point sources emitting a line with rest frequency $\nu_{\rm{em}}$ and luminosity function $\Phi(L)$.  If we divide our space into voxels with volume $V_{\rm{vox}}$, then the observed intensity in a given voxel is
\be
T = \frac{X_{LT}}{V_{\rm{vox}}}\sum_{i=1}^N L_i,
\ee
where $L_i$ are the luminosities of the $N$ galaxies contained within $V_{\rm{vox}}$ \citep{Lidz2011}. We have neglected beam effects here, i.e. a source contributes all of its intensity to the voxel it is contained within, and there is no smoothing effect spreading the intensity over multiple voxels. 

If we consider only voxels that contain exactly one emitter, the probability\footnote{For clarity, we will use the symbol $P$ when referring to power spectra and $\p$ when referring to probability distributions.} $\p_1(T)$ of observing intensity $T$ is given by
\be
\p_1(T) = \frac{V_{\rm{vox}}}{\overline{n}X_{LT}}\Phi\left(TV_{\rm{vox}}/X_{LT}\right),
\label{P1}
\ee
where 
\be
\overline{n}=\int_0^\infty \Phi(L)dL
\ee
is the average comoving number density of sources.  The probability of observing $T$ in a voxel with two sources is then
\be
\p_2(T) = \int\int \p_1(T') \p_1(T'')\delta_D(T-T'-T'')dT'dT'',
\ee
where $\delta_D$ is the Dirac delta function.  This simplifies to the convolution
\be
\p_2(T) = \int \p_1(T')\p_1(T-T')dT' = (\p_1\ast \p_1)(T).
\label{P2}
\ee
From this it is clear that the probability of observing $T$ in a pixel with $N$ sources is simply
\be
\p_N(T)=(\p_{N-1}\ast \p_1)(T).
\label{PN}
\ee
An empty pixel obviously will always give zero intensity, so $\p_0(T)=\delta_D(T)$.  

With Equations (\ref{P1}) and (\ref{PN}) we can recursively compute these probability distributions for pixels containing any arbitrary number of sources.  The full VID is then given by
\be
\p(T)=\sum_{N=0}^\infty \p_N(T)\Pn,
\label{PofT}
\ee
where $\Pn$ is the probability of observing a voxel that contains $N$ sources.  \citet{Lee2009} go on to replace the series of convolutions used to calculate $\p_N(T)$ with products in Fourier space.  However, as luminosity functions often span many orders of magnitude, it may be computationally easier to compute $\p_N(T)$ using convolutions.

If the sources are unclustered, then $\Pn$ is the Poisson distribution $\p_{\rm{Poiss}}(N,\Nbar)$ with mean $\Nbar=\overline{n}V_{\rm{vox}}$.  Equation (\ref{PofT}) can then be simplified further, as shown in \citet{Lee2009}.  However, since real galaxies trace the underlying dark matter distribution, we do see significant clustering which must be taken into account.  One method for $P(D)$ analysis of clustered sources is given by Equation (21) of \citet{Barcons1992}.  However, this method requires knowing all of the higher N-point statistics of the intensity distribution, which makes computing the VID intractable.  

We instead make use of the fact that the galaxy number-count distribution is known to be approximately lognormal \citep{Coles1991}.  Using this fact, we can follow \citet{Breysse2015} and assume that for each voxel there is an expectation value $\mu$ for the number of galaxies contained within it which depends on the value of the lognormal cosmic density field at that point.  The observed number of galaxies within that voxel will then be a Poisson draw from a distribution with mean $\mu$.  We can then write $\Pn$ as
\be
\Pn = \int_0^\infty \p_{LN}(\mu)\p_{\rm{Poiss}}(N,\mu)d\mu,
\label{PofN}
\ee
where $\p_{LN}$ is lognormal probability of finding a voxel with expectation value $\mu$.

The lognormal probability $\p_{LN}$ is computed assuming that the galaxy density field has density contrast in $\delta_{LN}(\vec{x})=[\mu(\vec{x})-\Nbar]/\Nbar$ in a voxel located at $\vec{x}$.  We can write $\delta_{LN}$ in terms of a Gaussian random variable $\delta_G$ as
\be
1+\delta_{LN}=\exp\left(\delta_G-\frac{\sigma_G^2}{2}\right),
\ee
where $\sigma_G$ is the variance of the Gaussian random field.  We can then write $\p_{LN}$ as
\be
\p_{LN}(\mu) = \frac{1}{\mu\sqrt{2\pi \sigma_G^2}}\exp\left\{-\frac{1}{2\sigma_G^2}\left[\ln\left(\frac{\mu}{\bar{N}}\right)+\frac{\sigma_G^2}{2}\right]^2\right\},
\ee
\citep{Kayo2001}.

The quantity $\sigma_G$ sets the overall ``strength" of the clustering, i.e. fields with a larger $\sigma_G$ have comparatively more voxels containing very many or very few sources, and comparatively fewer ``mid-range" voxels.  We can compute it from the power spectrum $P_G(k)$ of $\delta_G$ using
\be
\sigma_G=\int P_G(k) \left|W(\vec{k})\right|^2\frac{d^3\vec{k}}{(2\pi)^3},
\label{sg}
\ee
where $W(\vec{k})$ is the Fourier transform of the voxel window function.  The spectrum $P_G(k)$ can be calculated using the fact that the real-space correlation functions $\xi(r)$ of $\delta_{LN}$ and $\delta_G$ are related by
\be
\xi_G=\ln\left[1+\xi_{LN}(r)\right],
\label{xig}
\ee
\citep{Coles1991}, and the power spectra and correlation functions are related in the usual manner.  We thus only need to assume a power spectrum for our lognormal field to compute $\sigma_G$.  Here, we assume that this spectrum is given by $P_{\rm{LN}}(k)=\bar{b}^2P_m(k)$ calculated using Equations (\ref{Pk}) and (\ref{bmean}).  Note that since the bias $\bar{b}$ used here is luminosity weighted, this is slightly different from the power spectrum typically used for galaxy surveys.

In a realistic experiment, many of the fluctuations observed in a map will be caused by instrumental noise.  The measured intensity in a given voxel will then be the sum of the signal and noise contributions.  The noise will have its own VID $\p_{\rm{Noise}}(T)$ which is determined by the instrumental properties.  For example, in the case of simple Gaussian noise the noise VID is
\be
\p_{\rm{Noise}}(T)=\frac{1}{\sqrt{2\pi\sigma_N^2}}\exp\left(-\frac{T^2}{2\sigma_N^2}\right),
\ee
where the variance $\sigma_N$ is set by the survey sensitivity.  By the same arguments used in Equation (\ref{P2}), the VID for the sum of signal and noise is
\be
\p_{\rm{Total}}(T) = \left(\p_{\rm{Signal}} \ast \p_{\rm{Noise}}\right)(T).
\ee
Contributions from other sources of contamination, such as line or continuum foreground emission, can be added to the VID in a similar fashion.

To summarize the above formalism, the steps to compute a VID for a given model are as follows:
\begin{itemize}
\item Assume a luminosity function $\Phi(L)$, a voxel shape, and a cosmological model.
\item Compute the probability $\p_N(T)$ of observing $T$ in a voxel containing $N$ sources from $\Phi(L)$ using Equation (\ref{PN}).
\item Determine the mean number $\Nbar$ of galaxies/voxel using the assumed $\Phi(L)$ and voxel dimensions.
\item Compute a power spectrum for the lognormal galaxy field from the assumed cosmological model.
\item Calculate $\sigma_G$ from this power spectrum using Equations (\ref{sg}) and (\ref{xig}).
\item Use Equation (\ref{PofN}) to compute the probability $\Pn$ of observing a voxel with $N$ galaxies using the calculated $\Nbar$ and $\sigma_G$.
\item Sum $\p_N(T)\Pn$ over all values of $N$ as in Equation (\ref{PofT}).
\item Convolve the resulting VID with VIDs computed for instrumental noise and any foreground contamination.
\end{itemize}

Note that this method as presented here makes a subtle approximation about the halo bias.  By including the bias in our chosen $P_{LN}(k)$ spectrum, we take into account the fact that galaxies are more strongly clustered than the underlying dark matter.  However, putting it into the model in this manner effectively assigns each galaxy the average bias value, when in reality the brightest galaxies should be more strongly clustered.  The model therefore underestimates the number of very bright pixels.  Fortunately, we expect the effect of this to be small, especially given the immense uncertainties that currently exist in the modeling of $\Phi(L)$ for most intensity mapping lines.  For more discussion of bias, see Appendix \ref{bias}.
  
\section{Fiducial Line Emission Model}
The formalism described above can be readily applied to a wide variety of different lines and models.  In order to demonstrate its effectiveness, we will now without loss of generality apply the VID to a model of a CO intensity map.  The luminosity function of CO emission can be computed in a number of different ways (see, for example, \citet{Visbal2010}, \citet{Pullen2013}, \citet{Li2016}, \citet{Popping2016}), with different models often leading to wildly different forecasts \citep{Breysse2014}.  To make our example model as broadly applicable as possible, we model the CO luminosity function as a slightly modified Schechter function
\be
\frac{\Phi(L)}{(\textrm{Mpc}/h)^{-3}\ L_{\sun}}=\phi_*\left(\frac{L}{L_*}\right)^\alpha \exp\left(-\frac{L}{L_*}-\frac{L_{\rm{min}}}{L}\right),
\label{LFmodel}
\ee
where $p_i\equiv\left(\phi_*,\alpha,L_*,L_{\rm{min}}\right)$ are free parameters \citep{Schechter1976}.  This functional form should capture the essential features of a wide variety of models, with the high-luminosity cutoff caused by the reduced star formation efficiency in large galaxies and low-luminosity cutoff caused by the difficulty of forming galaxies in very low mass halos.  The quantity $L_{\rm{min}}$ here serves the purpose of the hard minimum mass or luminosity seen in most literature models, but replaced with an exponential cutoff both for added realism and to prevent numerical issues which can arise around hard cutoffs.

We choose values for our four parameters by fitting a Schechter function to the luminosity function plotted in Figure 8 of \citet{Li2016}.  This luminosity function is calculated from a suite of N-body simulations using the relation between halo mass and star formation rate from \citet{Behroozi2013}.  Star formation rates are then connected to CO luminosity through a series of empirical scaling relations \citep{Kennicutt1998,Carilli2013}.  The best fit Schechter function has parameters $\phi_*=2.8\times10^{-10}$, $\alpha=-1.87$, and $L_*=2.1\times10^6\ L_{\sun}$.  The Schechter function fits the \citet{Li2016} results reasonably well, though it does produce a steeper high-luminosity cutoff.  Since essentially nothing is known currently about CO emission from very faint galaxies, we somewhat arbitrarily choose $L_{\rm{min}}=5000\ L_{\rm{sun}}$.  

As for the bias, If we assume the linear relation between halo mass and CO luminosity from \citet{Pullen2013} Model A, this $L_{\rm{min}}$ corresponds to a halo mass of $2.5\e9$ solar masses.  This is comparable to literature values that usually place the CO luminosity cutoff around $10^9-10^{10}\ M_{\sun}$.  When computing the mean bias $\overline{b}$, we use this same linear mass-luminosity relation along with the Tinker form of the mass function and $b(M)$ \citep{Tinker2008,Tinker2010}.

\section{Results}
We will now forecast the constraining power of a COMAP-like experiment for our CO emission model.  The parameters of the full COMAP experiment can be found in Table 2 of \citet{Li2016}.  The planned survey would target the 115 GHz CO(1-0) line in 400 bands between $z=2.4$ and $z=2.8$, giving each band a width $\delta\nu = 10$ MHz.  The planned instrument has an angular resolution of 3 arcminutes with a total survey area of 6.25 deg$^2$.  Based on these parameters, we assume a model voxel which is a $3'\times3'$ square in the plane of the sky with a comoving depth set by the 10 MHz frequency bandwidth.  The survey aims for a noise/voxel of 5.8 $\mu$K, so we adopt a Gaussian noise VID with $\sigma_N=5.8\ \mu$K.

\begin{figure}
\centering
\includegraphics[width=\columnwidth]{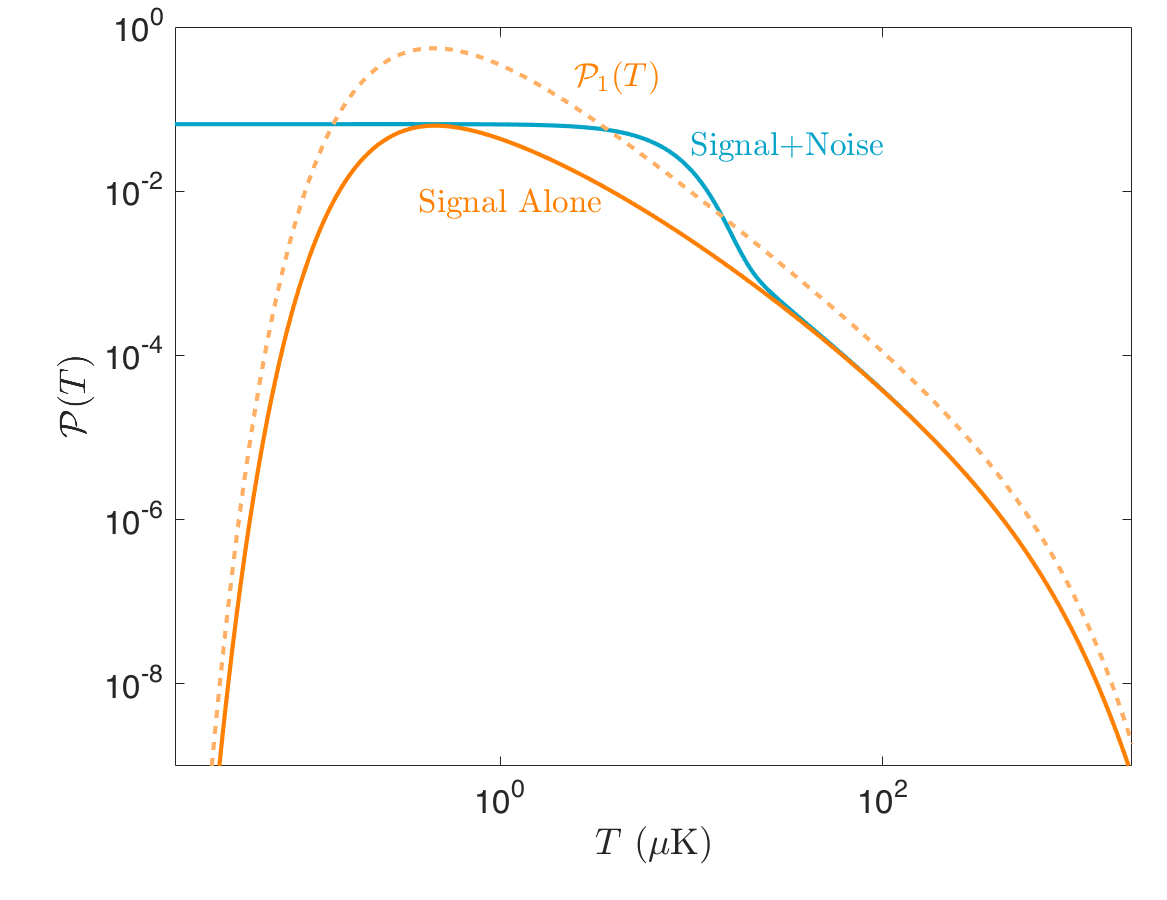}
\caption{VIDs for our fiducial model and experiment with (orange) and without (blue) instrumental noise.  Also plotted is the probability distribution $\p_1(T)$ for voxels that are known to contain exactly one source (dashed orange).  This distribution is simply a rescaling of our Schechter luminosity function.}
\label{VID}
\end{figure}

Figure \ref{VID} shows the VIDs for our fiducial model both with and without noise.  The dashed curve shows $\p_1(T)$, which is a simple rescaling of our fiducial Schechter function.  The amplitude of the VID is reduced from that of $\p_1(T)$ because the expected number of galaxies per voxel is only $\sim0.3$ in this setup, leading to a significant number of voxels that contain zero galaxies. This creates a delta function in $\p(T)$ at $T=0$ (not shown).  This suppression is less for brighter intensities though because some of the difference is made up by voxels that contain several sources.  These multiple-source voxels cause the VID to deviate a modest amount from the Schechter power law in the middle of the distribution.  The effect of the instrumental noise is to remap all of the faint voxels into a Gaussian distribution, with the signal VID dominating the bright end of the distribution.  These calculated VIDs agree well with simulated VIDs prepared using the method of \citet{Breysse2015}, as shown in detail in Appendix \ref{Sim}.

In order to study the constraining power of our example CO model, we will apply the Fisher matrix formalism.  This assumes that the likelihood distribution of the parameter values is a multivariate Gaussian centered on the fiducial values.  The Fisher matrix $F_{\mu\nu}$ for a model with parameters $p_\mu$ is given by
\be
F_{\mu\nu}=\sum_i\frac{1}{\sigma_i^2}\frac{\pa B_i}{\pa p_\mu}\frac{\pa B_i}{\pa p_\nu},
\label{Fisher}
\ee
where 
\be
B_i = N_{\rm{vox}}\int_{T_{\textrm{min},i}}^{T_{\textrm{max},i}}\p(T)dT
\ee
is the number of pixels in an intensity bin with edges $[T_{\textrm{min},i},T_{\textrm{max},i}]$ for a survey containing $N_{\rm{vox}}$ voxels \citep{Jungman1996a,Jungman1996b}.  The expected variance $\sigma_i^2$ on $B_i$ is assumed to be equal to $B_i$, i.e. we assume the bins obey Poisson statistics.  This should be a reasonable assumption for bins that contain many pixels, which are the bins that will contribute most of the signal-to-noise.  For simplicity, we use the VID computed for $z=2.6$, the central redshift of the survey, for all frequency bins.  The Fisher matrix computed from Equation (\ref{Fisher}) can then be inverted to obtain the covariance matrix of the model parameters.

We compute the Fisher matrix using five parameters: the four parameters of our model luminosity function and $\sigma_G$.  Adding $\sigma_G$ as a free parameter takes into account our uncertainty on both the average bias and the clustering behavior of the target galaxies.  Since our goal is to measure the luminosity function, we will report constraints on the four Schechter parameters marginalized over $\sigma_G$.  The result of this analysis for our fiducial model and experiment are shown in Figure 2.  The smaller orange ellipses show the constraints obtained for an ideal measurement with zero instrumental noise, with the only errors due to sample variance.  For an instrument with infinite sensitivity, these errors could be further reduced by observing a larger area of the sky.  The larger blue ellipses show the effect of the Gaussian COMAP-like instrumental noise.  Adding noise obviously makes the constraints somewhat worse, but there is still a substantial amount of constraining power.  The constraint on the low-luminosity cutoff $L_{\rm{min}}$ is hardest hit, as the cutoff occurs well below the intensity where noise begins to dominate.

\begin{figure}
\centering
\includegraphics[width=\columnwidth]{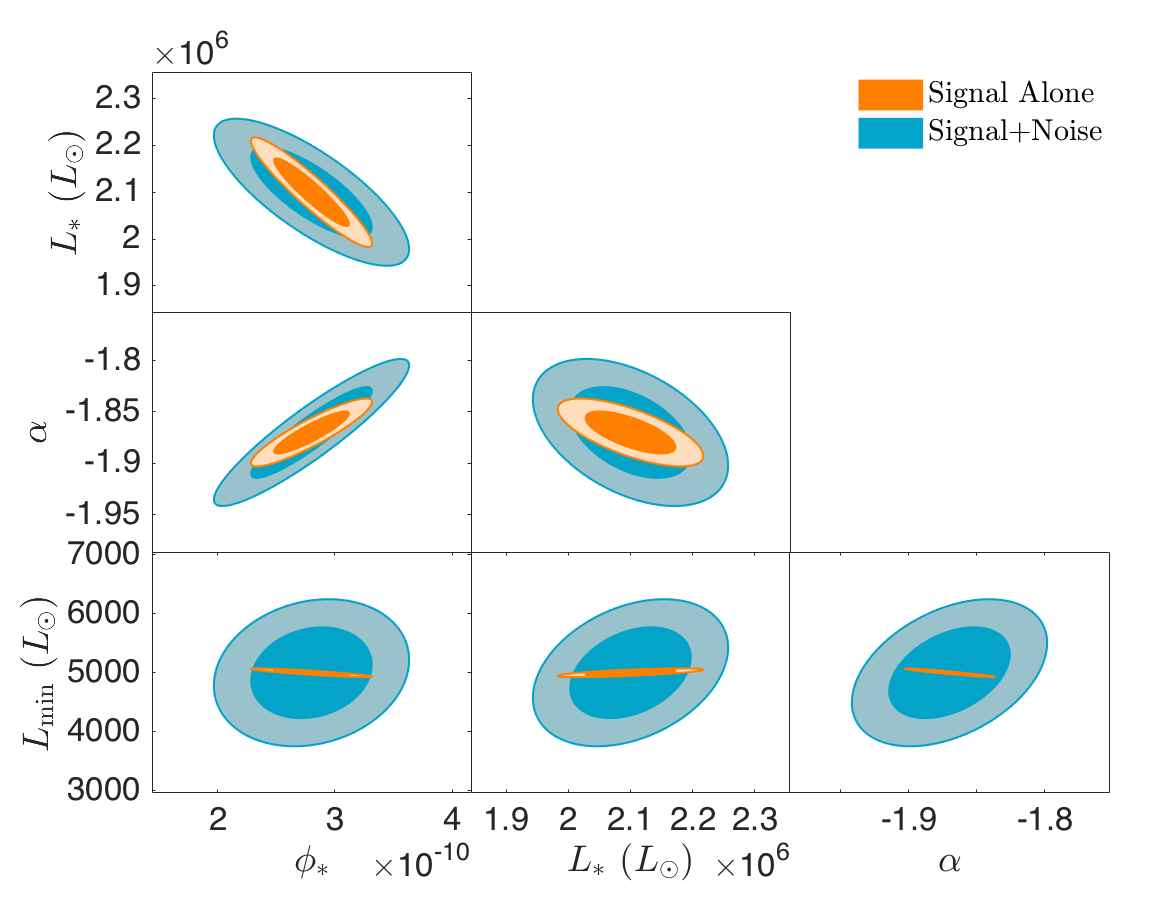}
\caption{Fisher constraints on the four parameters of our fiducial CO luminosity function both with (blue) and without (orange) Gaussian instrumental noise.  Dark ellipses show 1-$\sigma$ errors, light ellipses show 2-$\sigma$ errors.}
\label{Fiducial}
\end{figure}

\begin{figure}
\centering
\includegraphics[width=\columnwidth]{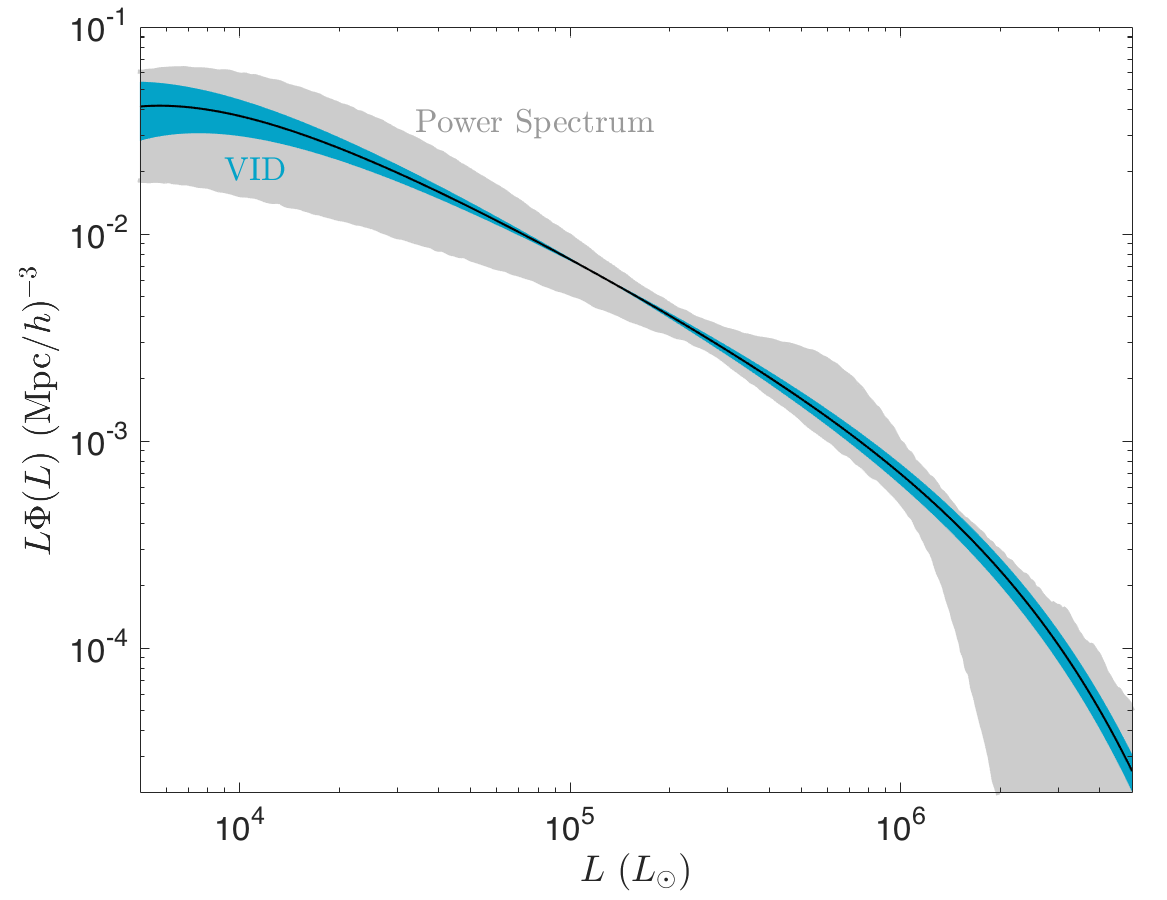}
\label{LFerrFig}
\caption{95\% confidence regions around our fiducial $\Phi(L)$ obtained from the above VID constraints including COMAP instrumental noise (blue) and from the COMAP power spectrum analysis presented in \citet{Li2016} (gray).}
\end{figure}

Using equation (\ref{LFmodel}), we can convert these constraints on the model parameters into constraints on $\Phi(L)$.  The resulting 95\% confidence region for the case with COMAP Full instrumental noise are shown in blue in Figure \ref{LFerrFig}.  We can compare these uncertainties to those forecasted in \citet{Li2016} using the COMAP power spectrum.  If we apply the same fractional uncertainties plotted in Figure 8 of \citet{Li2016} to our fiducial $\Phi(L)$, we get the 95\% confidence region plotted in grey.  Because the VID statistic is much better suited to measuring the luminosity function, it leads to substantially better constraints than the power spectrum despite using the same instrumental setup.  It should be noted that these constraints only hold if the real luminosity function has exactly the form given in equation (\ref{LFmodel}).  For this reason, computing uncertainties on $\Phi(L)$ in this manner likely underestimates the true error, especially at the faint end where the VID is noise dominated.

\section{Foreground effects}
The results shown above assume that the measured fluctuations in an intensity map are caused only by the source galaxies and instrumental noise.  However, as when studying the power spectrum, there are a number of effects that can alter the observed intensity fluctuations and reduce the constraining power of a given experiment.  Below we provide examples of how three of these effects, continuum foregrounds, line foregrounds, and gravitational lensing, can affect the constraints obtained from the VID.

\subsection{Continuum Foregrounds}
One of the most substantial difficulties facing any intensity mapping experiment is the presence of bright continuum foreground emission.  This problem is most evident in 21 cm experiments, where foreground contamination can be $\sim5$ orders of magnitude brighter than the signal \citep{Morales2010}.  Other lines face the same problems to a somewhat lesser degree.  For example, a COMAP like survey would observe at roughly 30 GHz, an area in frequency space familiar to CMB observers for containing large amounts of galactic synchrotron and free-free emission \citep{Planck2015a}, as well as contamination from radio point sources \citep{Keating2015}.  Indeed, the CMB itself creates a substantial amount of ``foreground" contamination at these frequencies.

Despite the overall strength of continuum foregrounds compared to the signal, they can be cleaned out of a map by taking advantage of their smooth frequency spectra.  The intensity mapping signal will have a significant amount of structure in frequency space, as it maps the distribution of galaxies along the line of sight.  Foregrounds such as synchrotron emission, however, are expected to be quite spectrally smooth.  This means that continuum contamination is confined to Fourier modes oriented near the plane of the sky, and it can be effectively removed by subtracting out these modes, as was demonstrated by \citet{Keating2015}.

The effects of this mode subtraction are straightforward to model in power spectrum space, as the only effect is to reduce the number of available Fourier modes.  Unfortunately, there is no clear way within our formalism to exactly replicate the effects of subtracting out only line-of-sight modes.  Studying this procedure accurately may require the use of simulated maps.  As an approximation to the correct foreground cleaning procedure, we consider a case where all of the $k=0$ modes are removed from a map, both along the line of sight and in the plane of the sky.  This will not exactly duplicate the true effect of continuum foregrounds, but it provides a rough estimate of the significance of the effect.  For simplicity, we will also consider a spatially uniform foreground

Consider then a map made up of three contributions: the usual CO signal, Gaussian, zero-mean instrumental noise, and a spatially and spectrally smooth foreground with intensity $\overline{T}_{CF}$.  The observed intensity in any given voxel will then be
\be
T_{\rm{obs}}=\overline{T}_{\rm{CO}}+\Delta T_{\rm{CO}}+\Delta T_{\rm{Noise}}+\overline{T}_{CF},
\ee
where we have divided the CO signal into mean and fluctuation parts.  Subtracting out the $k=0$ modes means that we can only observe
\be
\Delta T_{\rm{obs}}=T_{\rm{obs}}-\overline{T}=\Delta T_{\rm{CO}}+\Delta T_{\rm{Noise}},
\ee
where $\overline{T}=\overline{T}_{\rm{CO}}+\overline{T}_{CF}$.  The probability of observing a voxel with fluctuation $\Delta T$ is then
\be
\p_\Delta(\Delta T)=\p(\Delta T+\overline{T})
\label{Pdelta},
\ee 
where $\p(T)$ is the original signal+noise VID.

When computing the Fisher matrix for this new model, the number of voxels in a fluctuation-space bin is
\be
B_i=N_{\rm{vox}}\int_{\Delta T_{\textrm{min},i}}^{\Delta T_{\textrm{max},i}}\p_\Delta(\Delta T)d\Delta T.
\ee
Using Equation (\ref{Pdelta}) we can rewrite this as
\be
B_i=N_{\rm{vox}}\int_{\Delta T_{\textrm{min},i}+\overline{T}}^{\Delta T_{\textrm{max},i}+\overline{T}}P(T)dT.
\ee
We have now introduced a sixth unknown parameter $\overline{T}$ into our calculation.  In the presence of our simple foreground, it is no longer known \emph{a priori} which absolute intensity $T$ corresponds to a given fluctuation $\Delta T$.  In other words, we can only measure the VID up to an additive constant in every voxel.  Fortunately, we can easily add this extra parameter to our Fisher analysis to determine how it affects our constraints.

\begin{figure}
\centering
\includegraphics[width=\columnwidth]{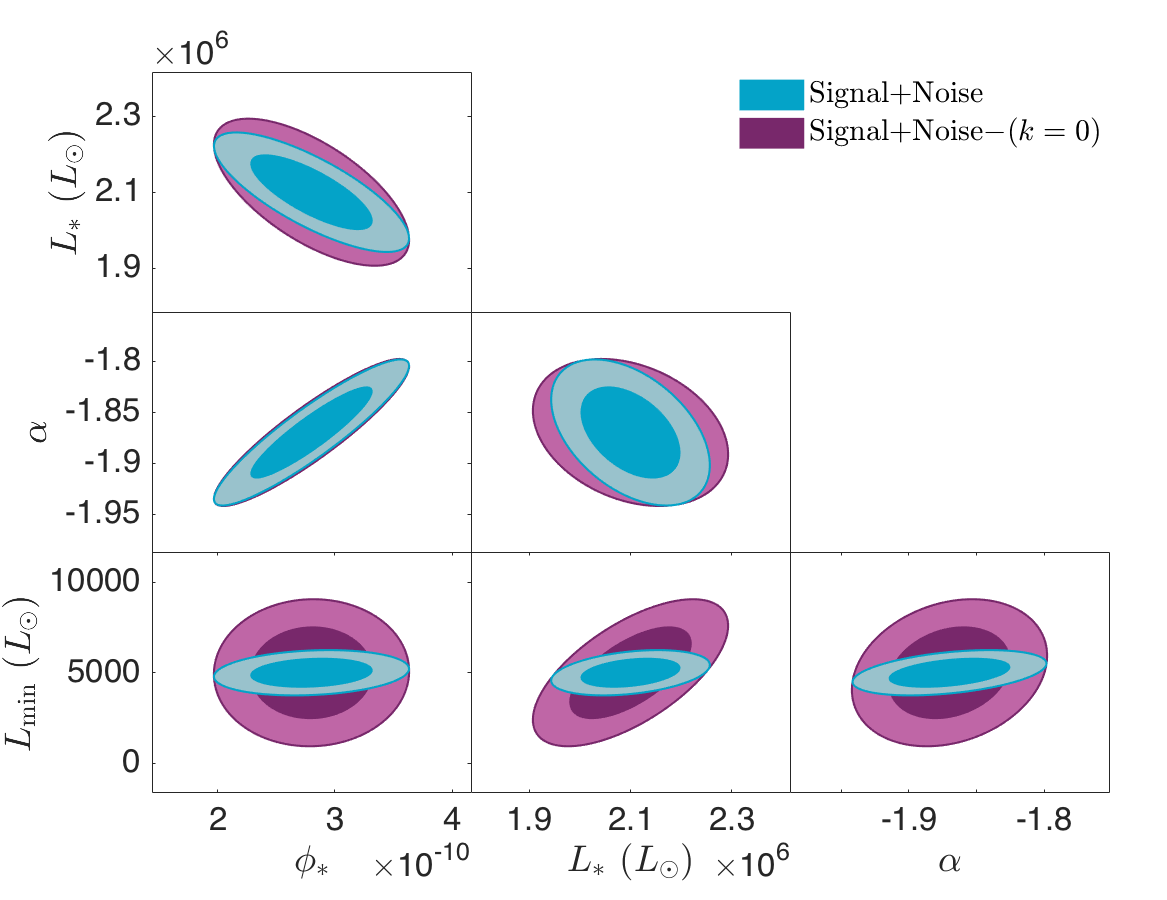}
\caption{Fisher constraints for a CO map where all $k=0$ modes have been subtracted to remove continuum (purple) compared with the signal+noise constraints from Figure \ref{Fiducial}.}
\label{CF}
\end{figure}

Figure \ref{CF} shows the effect of this added uncertainty, with the signal+COMAP noise constraints from Section 5 shown in blue and the new continuum-subtracted constraints shown in purple.  The continuum-subtracted constraints are marginalized over the unknown mean intensity $\overline{T}$ in addition to $\sigma_G$.  The foreground subtraction procedure worsens the constraints slightly, but the effect is not dramatic.  The effect of this subtraction on the $\Phi(L)$ constraints shown in Figure \ref{LFerrFig} are given in Appendix \ref{LFerrApp}.  Simply subtracting out the mean of the map, therefore, does not lead to a substantial loss of constraining power.  Note that a true spatially-varying foreground may leave some residual contamination which will affect these constraints.  This effect will be highly model-dependent, so we leave it for future work.

\subsection{Line foregrounds}
Intensity maps can also suffer from interloper emission from other spectral lines emitted by galaxy populations at different redshifts.  21 cm surveys are not expected to see significant line contamination, as there are few other bright lines at such low frequencies \citep{Gong2011}.  Surveys aiming for other lines, however, could see interloper lines bright enough to rival or dominate over the target lines \citep{Gong2014,Breysse2015,Silva2015}.  In such cases, the ability to study the original target line can be significantly degraded.

The literature contains several proposed means of recovering the power spectrum of a target line from a foreground-contaminated map.  If one has access to other data in the target volume, either traditional data like a galaxy map or an intensity map of a different line, it is possible to cross-correlate the two data sets together to isolate emission from a single redshift \citep{Visbal2011}.  It may also be possible to use anisotropies in the power spectrum to separate signal from foreground  \citep{Lidz2016,Cheng2016}.  One could also seek simply to mask out voxels where the foreground contribution is brightest, either blindly \citep{Breysse2015} or by using galaxy surveys to locate bright foreground emitters \citep{Crites2014}.

We can divide line contamination into two broad categories based on how it affects the power spectrum.  The most obviously problematic lines are those for which the mean foreground intensity $\overline{T}_F$ is greater than $\overline{T}_{\rm{CO}}$.  We will refer to these lines as ``clustering foregrounds", as the foreground dominates over the signal in the clustering component of the power spectrum.  Examples of clustering foregrounds are H$\alpha$, OIII, and OII in Ly$\alpha$ surveys \citep{Gong2014} and higher-order CO rotational lines in CII surveys \citep{Silva2015}.  It is also possible for foreground lines that are substantially fainter than the signal on average to produce a small population of very bright sources that contribute a disproportionate amount of shot noise to a map, leading us to refer to them as ``shot-noise foregrounds".  The model considered in \citet{Breysse2015} for HCN contamination in CO surveys is an example of such a foreground.

Here we seek to understand how contamination from foreground lines affects the VID statistic.  For the sake of simplicity, we will continue using our fiducial CO signal model and invoke hypothetical foregrounds to test their effects.  We leave for future work a detailed exploration of signal and foreground models for surveys targeting CII, Ly$\alpha$, and other lines.  We choose the luminosity functions of our fiducial foreground lines to best demonstrate the behavior of the two types of foregrounds described above.  

The first line, which we will name FG1, we choose to be a shot-noise foreground with $\overline{T}_{\rm{FG1}}=0.1\overline{T}_{\rm{CO}}$ and $P_{\rm{shot,FG1}}=2P_{\rm{shot,CO}}$.  This roughly duplicates the behavior of the HCN model in \citet{Breysse2015}, and is something of a worst-case scenario for a CO intensity mapping survey.  The second line, which we will name FG2, we choose to be a clustering foreground with $\overline{T}_{\rm{FG2}}=5\overline{T}_{\rm{CO}}$ and $P_{\rm{shot,FG2}}=25P_{\rm{shot,CO}}$.  Though no lines of this type are expected to appear in CO surveys, this is the type of line expected to cause issues in high-redshift CII and Ly$\alpha$ surveys.  We assign both FG1 and FG2 an emission frequency of 88 GHz, which is the rest frequency of HCN(1-0).  We then choose values for the four Schechter parameters of each line to reproduce the desired power spectra.  The chosen parameter values can be found in Table \ref{FGpar}.

\begin{table}
\centering
\caption{Luminosity function parameters for fiducial CO model and hypothetical foreground lines}
\begin{tabular}{ccccc}
\hline
Line & $\phi_*$ & $L_*$ ($L_{\sun}$) & $\alpha$ & $L_{\rm{min}}$ ($L_{\sun}$) \\
\hline
CO(1-0) & $2.8\e{-10}$ & $2.1\e6$ & $-1.87$ & 5000 \\
FG1 & $4.1\e{-18}$ & $6.5\e8$ & $-2.26$ & 500 \\
FG2 & $ 5.1\e{-10}$ & $3.4\e6$ & $-1.6$ & 5000 \\
\hline
\end{tabular}
\label{FGpar}
\end{table}

\begin{figure*}
\centering
\includegraphics[width=\textwidth]{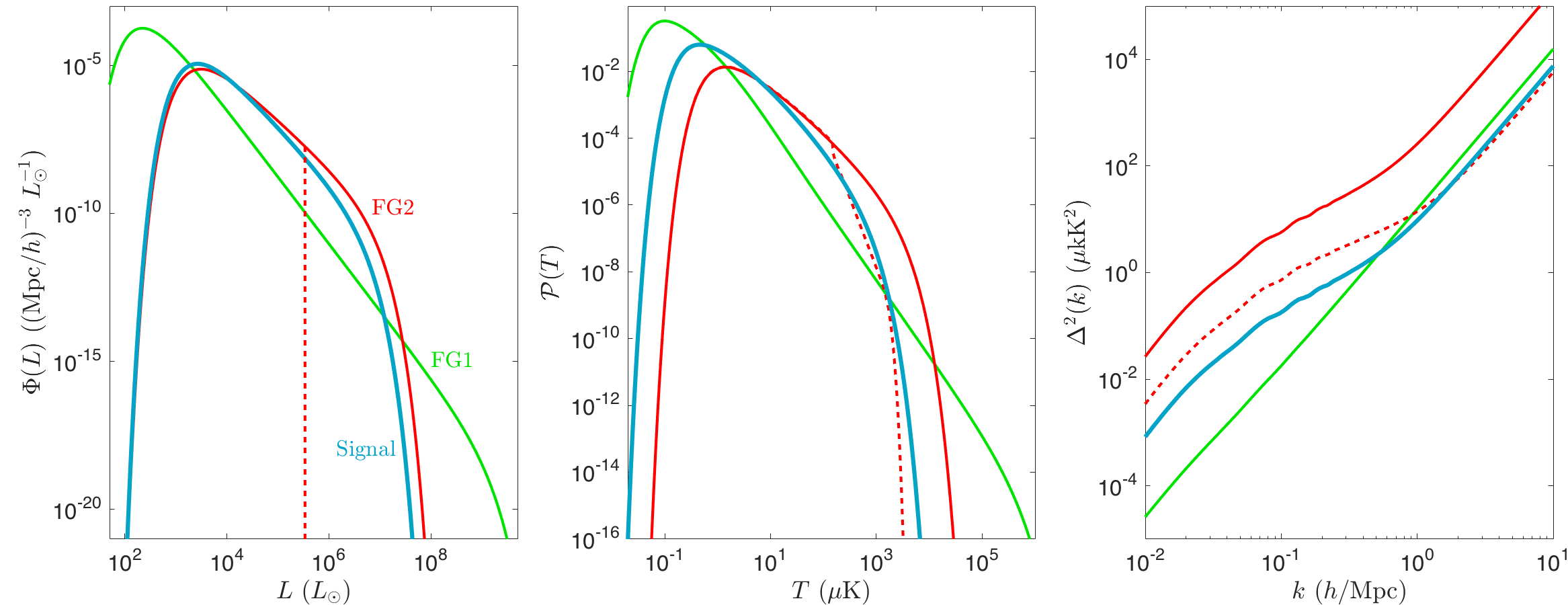}
\caption{Luminosity functions (left), VIDs (center), and power spectra (right) for the fiducial CO signal (blue), shot noise foreground FG1 (green), and clustering foreground FG2 (red).  Dashed red curves show the effect of masking FG2 emitters brighter than $L_{*,\rm{FG2}}/10$.}
\label{FGs}
\end{figure*}

Figure \ref{FGs} shows the luminosity functions, VIDs, and power spectra of the fiducial CO signal compared with those of the FG1 and FG2 models.  Power spectra are computed using matter power spectra from CAMB \citep{Lewis2011}.  The different behaviors of the two types of foregrounds can be clearly seen.  The high-luminosity tail of FG1 leads to a shot-noise dominated power spectrum, which is strong enough to compete with the signal despite the overall weakness of the FG1 line.  The FG2 luminosity function is very similar to that of the CO signal, but contains somewhat more bright sources.  Since these sources are also closer to the observer, this leads to a VID that dominates over that of CO at most observed intensities, and a power spectrum that dominates on all scales.  Note that the power spectra plotted here are unprojected, which means that the foreground spectra would be amplified even more relative to the signal in a true measurement.

It should be noted that the parameters in Table \ref{FGpar} do not uniquely determine the power spectra shown in the right-hand panel of Figure \ref{FGs}.  Because we have four free luminosity function parameters from which to determine the two terms of the power spectrum, we could choose an infinite number of different parameters and generate the exact same power spectra.  The particular four parameters shown here have thus been chosen somewhat arbitrarily.  As we are only using these fictional lines for a proof of concept, the exact parameter choices are not particularly important.  However, this does serve to again illustrate the limitations of power spectra, as the same measured spectrum could be the result of many different luminosity functions.

\begin{figure}
\centering
\includegraphics[width=\columnwidth]{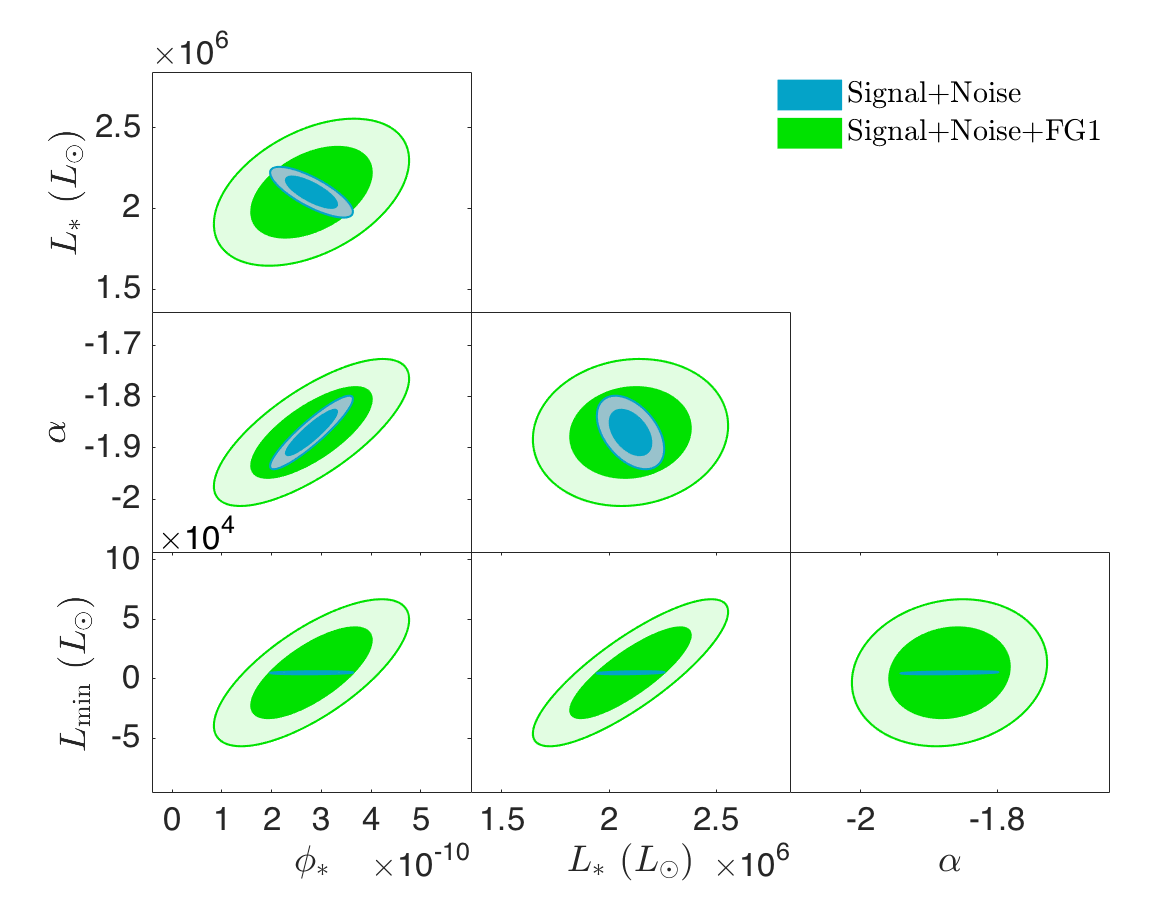}
\caption{Fisher constraints on the CO signal parameters marginalized over the parameters of the shot-noise foreground FG1 (green), compared with the original CO constraints (blue).}
\label{FG1}
\end{figure}

Figure \ref{FG1} shows the effect of including the shot noise foreground model in our Fisher analysis.  We allow the parameters of the foreground model to vary along with those of the CO signal and marginalize over the foreground luminosity function parameters as well as the $\sigma_G$ values for both lines.  The result is that the constraining power diminishes, but not prohibitively.  
With the exception of the low-luminosity cutoffs, which mostly affect the noise-dominated portion of the VID, the signal and foreground parameters are only weakly degenerate, and the constraints are fairly good.  This is a marked improvement over what is seen in power spectrum space, where the shot noise components of the signal and foreground spectra are exactly degenerate.  The constraints on $\Phi(L)$ with this foreground component included are given in Appendix \ref{LFerrApp}.

\begin{figure}
\centering
\includegraphics[width=\columnwidth]{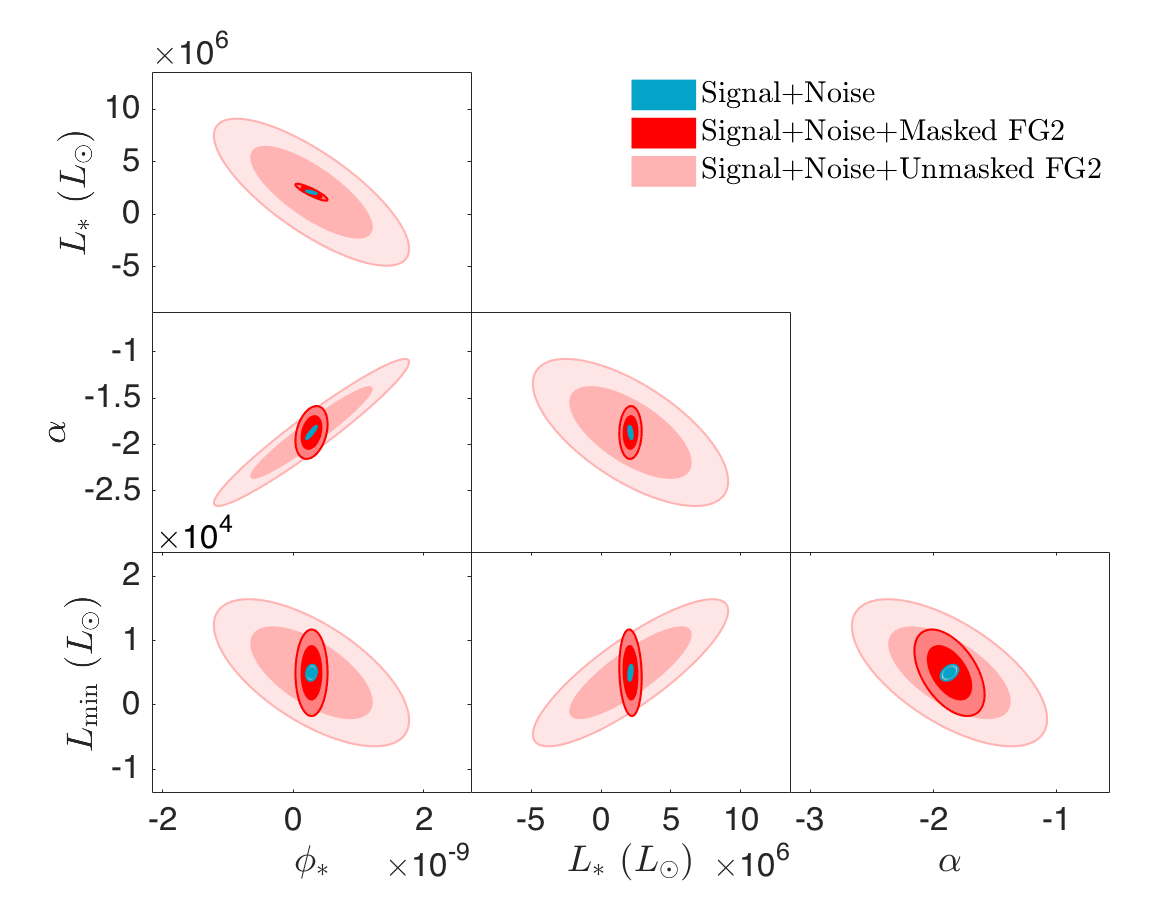}
\caption{Fisher constraints on the CO signal parameters marginalized over the parameters of the clustering foreground FG2 (light red), compared with the original CO constraints (blue).  Constraints with FG2 emitters brighter than $L_{*,\rm{FG2}}$ masked out are shown in red.}
\label{FG2}
\end{figure}

Figure \ref{FG2} demonstrates that, as one might expect, the effect of the clustering foreground FG2 is much more dramatic.  Marginalizing over the foreground parameters and the two $\sigma_G$ values leaves significantly worse constraints on the CO model.  In this case, all of the parameters except $\alpha$ have fractional uncertainties greater than unity.  This is caused by the fact that the combined VID cannot easily distinguish the two similar luminosity functions, leading to significant degeneracies.  Because these constraints are so poor, we can do little more than set upper limits on the target luminosity function.

However, we can improve on this result significantly by masking out bright foreground voxels \citep{Gong2013,Breysse2015}.  Typically this is done through the use of some proxy observable which is correlated with foreground luminosity, but we will simply dictate that all sources with FG2 luminosity greater than $L_{*,\rm{FG2}}/10$ are masked out of the map.  This corresponds to a few percent of the total number of voxels.  The effects of this masking on the luminosity function, VID, and power spectrum are shown by the dashed red curves in Figure \ref{FGs}.  Though this amount of masking still leaves a significant FG2 power spectrum, we can see from the dark red ellipses in Figure \ref{FG2} that the constraints from the VID have been substantially improved.  The resulting luminosity function constraints can be seen in Appendix \ref{LFerrApp}.

The exact value of the masking luminosity in a real survey will depend on which observables are used to find foreground emitters.  Our choice of $L_{*,\rm{FG2}}/10$ is somewhat arbitrary given that the CO line is not expected to suffer from issues with clustering foregrounds.  However, the results given here should be taken as fairly pessimistic, as the mask we have used is insufficient to reduce the FG2 power spectrum below that of CO.  

\subsection{Gravitational Lensing}
In addition to emission from other astrophysical sources, structure between the observer and the emitters can affect the VID through gravitational lensing.  Lensed galaxies have their positions on the sky altered slightly, their shapes distorted, and their intensities magnified.  Since the voxels in an intensity map are large compared to any single galaxy, only the latter effect is important to consider when computing the VID.  Lensing magnification, both weak and strong, will change the apparent luminosity function of the source population, which in turn will alter the VID in ways that could systematically affect luminosity function constraints.  Lensing effects are expected to increase substantially with redshift, so in addition to our usual $z\sim3$ CO model we will consider a case where we use the same luminosity function but take the emission redshift out to $z=7$.

To estimate the effect of lensing on the VID, we need to compute the probability $\pmag(m)$ of a given galaxy to have a magnification between $m$ and $m+dm$.  The observed luminosity after magnification is $L'=mL$.  $\pmag(m)$ contains contributions from both the large-scale matter distribution as well as compact, virialised halos.  To estimate the former, we adopt the method of \citet{Das2006}.  First, we divide the intervening mass distribution up to the target redshift $z$ into $N$ uncorrelated, thin mass sheets.  Each sheet $i$ spans a comoving radius between $r_i$ and $r_i+1$ and has central redshift $z_i$.  For each sheet, we consider fluctuations in the projected surface mass density $\Sigma_i$, where we have defined the surface density constrast
\be
x(\thb,z_i)\equiv\frac{\Sigma(\thb,z_i)-\overline{\Sigma}(z_i)}{\overline{\Sigma}(z_i)},
\ee
as a function of the sky position $\thb$.  In the Limber approximation \citep{Limber1953,Rubin1954}, the two-dimensional power spectrum $P_2(\ell,z_i)$ for $x(\thb,z_i)$ is given by
\be
P_2(\ell,z_i)=\frac{1}{\ell(r_{i+1}-r_i)^2}\int_{\ell/r_{i+1}}^{\ell/r_i}P_{NL}(k,z_i)dk,
\ee
where $\ell$ is the magnitude of the two-dimensional Fourier wavenumber and $P_{NL}(k,z_i)$ is the nonlinear three-dimensional matter power spectrum evaluated using \texttt{Halofit} \citep{Smith2003}.  The rms fluctuation $\sigma_2^2$ smoothed over an angular scale $\theta_0$ is then
\be
\sigma_2^2(\theta_0,z_i)=\frac{1}{2\pi}\int \ell P_2(\ell,z_i) e^{-\ell^2\theta_0^2} d\ell.
\ee

In \citet{Das2006}, the following parametric non-Gaussian PDF $\p_x(x)$ for the density fluctuation $x$ is found to be a good fit to numerical simulations:
\be
\p_x(x)=\frac{N}{x}\exp\left[-\frac{(\ln(x)+\omega^2/2)^2(1+A/x)}{2\omega^2}\right].
\ee
This is essentially a one-parameter family, as the three parameters $N$, $A$, and $\omega^2$ are fixed by $\sigma_2^2$ through the requirements that $\left<x\right>$=1, $\left<x^2\right>-\left<x\right>^2=\sigma_2^2$, and that the PDF is normalized to unity.  

In the weak lensing regime, the convergence $\kappa$ receives a contribution 
\be
\kappa_i(\thb)=\frac{\Sigma(\thb,z_i)-\overline{\Sigma}(z_i)}{\Sigma_c(z_i)},
\ee
from each mass sheet.  We have introduced the critical surface density
\be
\Sigma_c(z_i)=\frac{c^2}{4\pi G}\frac{D_s}{D_l(z_i)\left[D_s-D_l(z_i)\right]},
\ee
where $G$ is Newton's gravitational constant, $D_s$ is the comoving radial distance to the target redshift, and $D_l$ is that to the lens-plane redshift $z_i$.  An overdense patch therefore contributes $\kappa>0$ and an underdense patch contributes $\kappa_i<0$.

A distribution for the cumulative convergence $\kappa=\sum_i\kappa_i$ can be estimated numerically by randomly drawing $x$ for each mass sheet according to the corresponding PDF $\p_x(x)$.  This gives the convergence PDF $\p_\kappa(\kappa)$. Using the weak-lensing relation $m=(1-\kappa)^{-2}$, we can then compute the source-plane magnification PDF 
\be
\pmag(m)=\frac{(1-\kappa)^5}{2}\p_{\kappa}(1-m^{-1/2}),
\label{pmweak}
\ee
which accounts for the non-Gaussian statistics of weak (de-)magnification fairly well.

\begin{figure}
\centering
\includegraphics[width=\columnwidth]{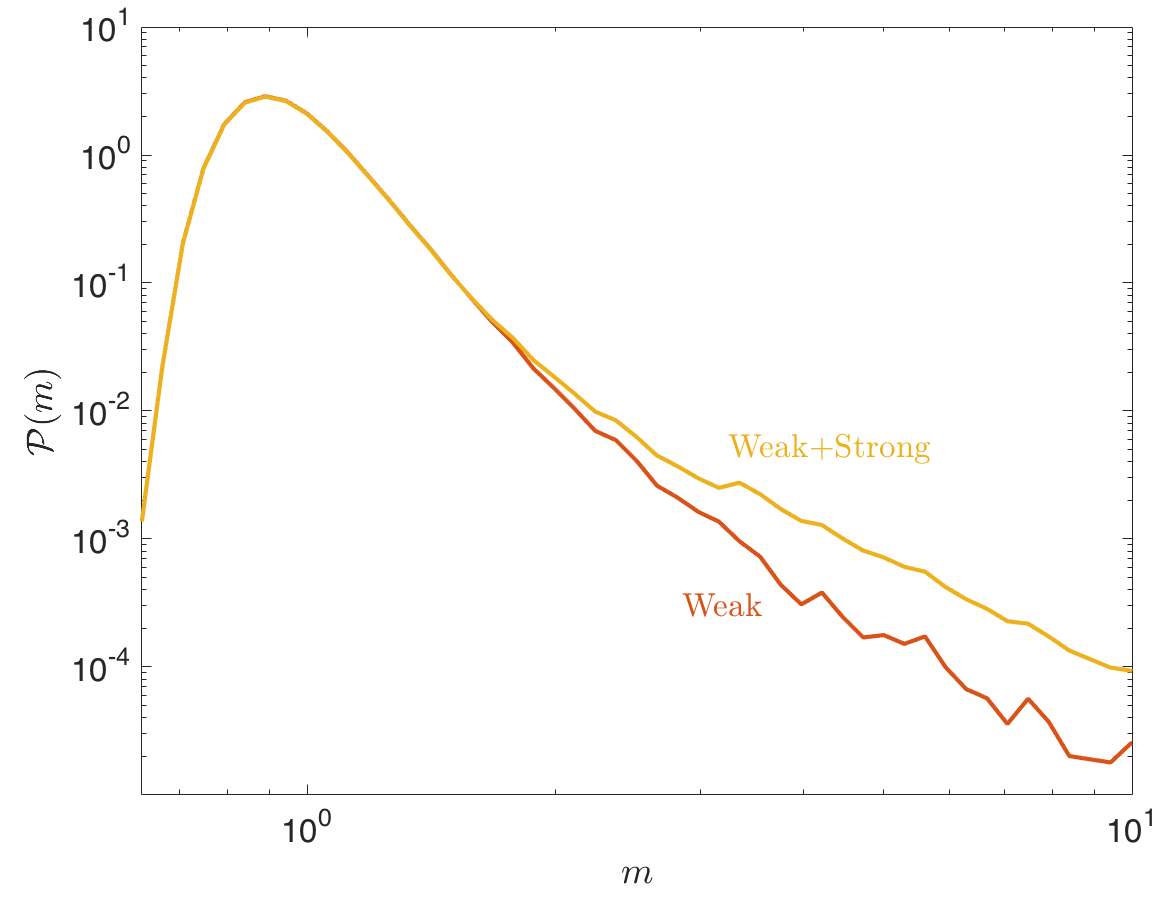}
\caption{Magnification PDFs drawn both with (yellow) and without (red) the power-law strong-lensing tail.}
\label{pmag}
\end{figure}

However, the method of \citet{Das2006} underestimates the effect of strong magnification from virialised lenses.  In particular, the large-$m$ tail is expected to decay roughly as $m^{-3}$ \citep{Takahashi2011}.  This can be remedied by manually adding a power law tail to Equation (\ref{pmweak}),
\begin{multline}
\pmag(m)\rightarrow \pmag(m)+\Theta(m-1)\exp\left[\frac{1}{4(m-1)^4}\right] \\ \times\frac{(1-\kappa_0)^3}{2}\p_{\kappa}(\kappa_0)\left(\frac{m}{m_0}\right)^{-3},
\label{pm}
\end{multline}
where a typical matching point is $\mu_0=(1-\kappa_0)^{-2}\sim3$ and $\Theta$ is the Heaviside function.  The resultant semi-analytic model is found to agree reasonably well with ray tracing of N-body simulations.  Figure \ref{pmag} shows magnification PDFs both with and without strong lensing for a given draw from $\p_x$ with source redshift $z=7$.  The two PDFs are similar in the low-magnification regime, but differ substantially at higher magnifications.

\begin{figure*}
\centering
\includegraphics[width=\textwidth]{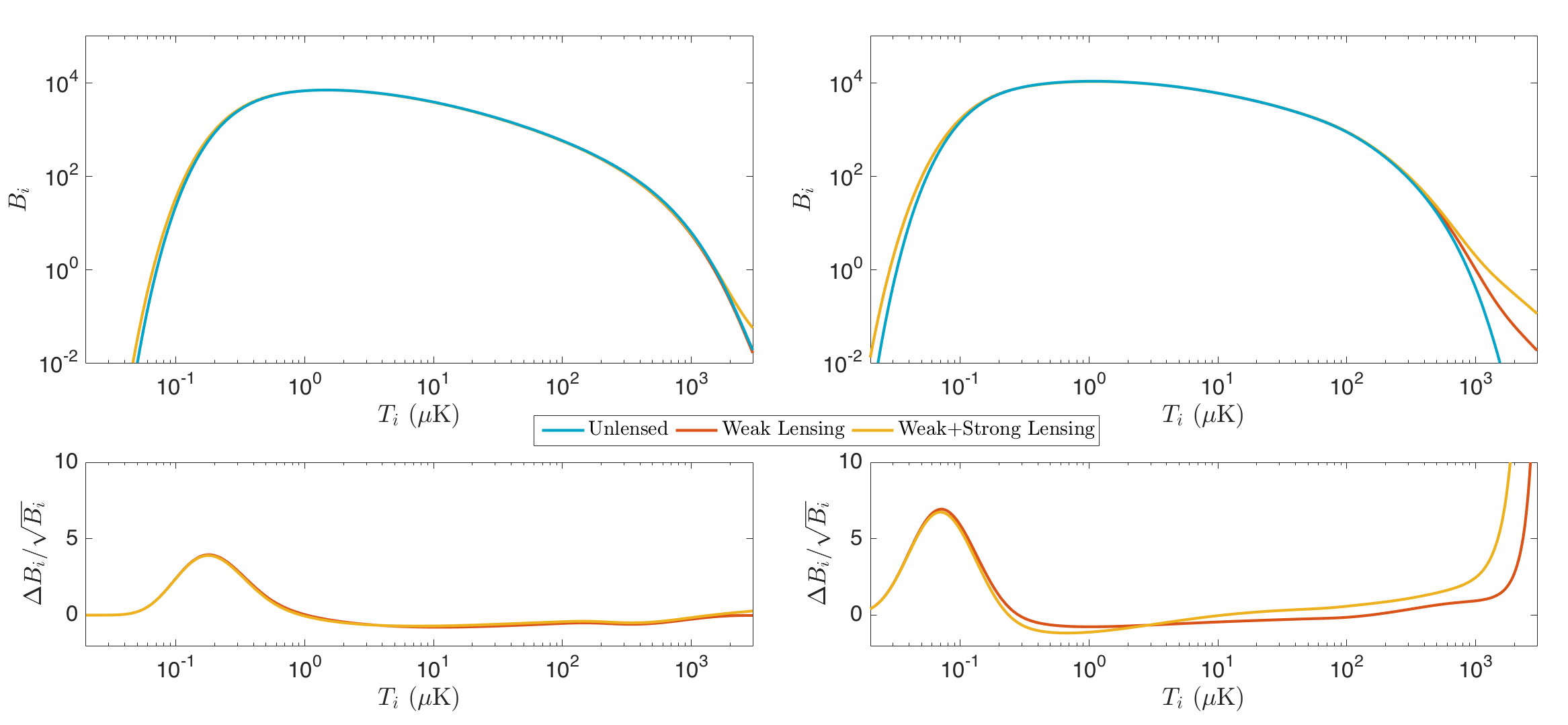}
\caption{Top row: Number of pixels in bins of width $\Delta\log T=0.012$ dex at redshift $z=3$ (left column) and $z=7$ (right column) for our unlensed fiducial CO model (blue), a weakly lensed model (red), and a full weak+strong lensed model (yellow).  Bottom row: Difference between the lensed and unlensed voxel counts divided by the Poisson errors used in our Fisher analysis.}
\label{lens}
\end{figure*}

The effect of magnification is to alter the apparent luminosity function of the source galaxies, replacing it with
\be
\Phi'(L)=\int \frac{\pmag(m)}{m}\Phi\left(\frac{L}{m}\right)dm,
\ee
as galaxies appear brighter or fainter due to lensing.  If we compute the VID from $\Phi'(L)$, we see that the distribution is altered somewhat, as shown in Figure \ref{lens}.  We have plotted both the VIDs for the full $\pmag$ as well as the weak-lensing-only $\pmag$ at both $z=3$ and $z=7$.  Note that in order to better understand the size of the effect, we have plotted these VIDs as bin counts rather than $\p(T)$.  If we divide the difference between the lensed and unlensed bin counts by the square root of the unlensed counts, we can get an idea of how strong the lensing effect is relative to the Poisson error in a given bin.  The effect is visible at both redshifts, and as expected it increases as redshift increases.  The primary impact is to make the cutoffs at both ends less sharp.  This will thus likely not be a hugely significant effect for near-future experiments, as the low-luminosity cutoff is below the noise limit and there are relatively few pixels above the high-luminosity cutoff.  However, future experiments with sufficient sensitivity and area may need to take into account magnification effects when attempting to constrain luminosity functions, especially when targeting high redshifts.

\section{Discussion}
The constraining power of the VID statistic is clear from the above results.  These constraints cannot be obtained using the power spectrum alone, as the power spectrum only measures the quantities $P_{\rm{shot}}$ and $\overline{T}^2\overline{b}^2$.  This means that the power spectrum can only constrain at most two of the four model parameters of our luminosity function model, resulting in significant degeneracies.  The only way to obtain useful results for a model with more than two parameters is then to apply priors.  However, as shown in Figure 2, the VID allows us to constrain our four-parameter Schechter model to order $\sim10\%$ even without prior data.  Even the low-end exponential cutoff $L_{\rm{min}}$, which falls well below our noise limit, can be constrained by the VID measurement.  

This effect can be clearly seen in Figure \ref{LFerrFig}, where the VID produces much better constraints on $\Phi(L)$ than the power spectrum model from \citet{Li2016}.  The constraints are improved significantly despite the fact that both forecasts use the same experimental setup and very similar luminosity functions.  As mentioned above, however, the constraints plotted in \ref{LFerrFig} likely somewhat underestimate the luminosity function errors, as both the VID and power spectrum constraints assume that the CO emitters have a luminosity function exactly described by a single model.  More realistic constraints could be obtained, for example, by using a spline model as in \citet{Glenn2010}, or by using the values of $\Phi(L)$ in different luminosity bins as the model parameters.

Though the VID provides considerably more information about the luminosity function, the power spectrum likewise contains information that is not present in the VID.  Just as the power spectrum contains only integrals over the luminosity function, the VID contains only integrals over the galaxy power spectrum.  The VID thus leaves out almost all of the information about the spatial distribution of galaxies.  This is intuitively obvious, since a one-point statistic by definition does not take into account voxel locations.  Because of this, the VID will not be nearly as effective as the power spectrum when attempting to measure, for example, baryon acoustic oscillations.  The VID and the power spectrum are highly complementary statistics, and using both to study a map will yield substantially more information than using only one or the other.

There are some subtleties involved when attempting to directly combine these two statistics, however.  Though they contain different information, there is a substantial amount of covariance between the VID and the power spectrum of a map.  For example, observing a region of space with a greater-than-average number of very bright sources will obviously yield more very bright voxels, but will also yield an excess amount of shot noise power.  Since shot noise is scale-independent, the net effect of this will be to create covariance between the size of the brightest voxel bins and the amplitude of the power spectrum on all scales.  Accurately treating covariances like this will like require detailed numerical simulations, so we leave a full study of this issue for future work (Breysse \& Li, in prep.).

As expected, the various forms of foreground contamination we studied do add extra uncertainty to our forecasts.  However, the VID retains significant constraining power even in the face of this contamination.  Our model of a continuum-subtracted map, though simplified, suggests that the process of cleaning foregrounds like Galactic dust and synchrotron will not destroy our ability to constrain the luminosity function from the VID.  This is due to the fact that the continuum foreground in our model merely adds a constant to every voxel, and this additive shift is not degenerate with any of our Schechter parameters.  The real situation, with foregrounds that vary in amplitude along different lines of sight, will be somewhat more complicated, but we do not expect the end result to be radically different than that plotted in Figure \ref{CF}.

The VID statistic also performs fairly well in the presence of line foregrounds.  Even without any attempts to clean interloper lines out of the map, the parameters $L_*$, $\alpha$, and $\phi_*$ are well constrained in the presence of a shot noise foreground, though degeneracies with the foreground line reduces our ability to constrain $L_{\rm{min}}$.  This scenario is something of a worst-case for a CO intensity map, so it is encouraging that the method remains viable.  If an analysis of such a map relied solely on the power spectrum, it would be essentially unable to use the shot noise information at all since the signal and foreground would be perfectly degenerate, so the value of the VID is clear.  The unmasked clustering foreground is less optimistic looking, though we are still able to obtain useful constraints on the power-law slope of the luminosity function.  However, surveys that suffer from this type of foreground, such as those targeting CII and Ly$\alpha$, already plan to mask out foreground sources \citep{Crites2014}.  Even with conservative enough masking that FG2 still dominates the power spectrum, the VID produces strong constraints on everything except $L_{\rm{min}}$.  The results would improve further with more aggressive masking or the application of other foreground-cleaning methods.

As for gravitational lensing, which can be thought of as contamination from foreground mass, the effect is very weak for our fiducial $z\sim3$ model and only slightly stronger when we take our emission out to $z\sim7$.  As shown in Figure \ref{lens}, the biggest effects are on the two exponential tails of the VID.  On the faint end, we see an excess due to weak lensing along underdense lines of sight.  Bins in this region are shifted by several sigma, but this effect will be negligible in a real observation because these bins will be strongly noise dominated.  As for the bright end, there is an excess, mostly caused by strongly lensed sources.  This could yield underestimates of the bright end slope of the luminosity function, but there are few enough voxels in this part of the distribution that the effect will likely not be large, at least for early surveys.  However, as surveys become more sensitive and target larger volumes, this effect will only become more important.  It is important to note though that gravitational lensing is somewhat different than the other types of contamination in that the observed luminosity function can in principle be deconvolved after the fact.  This could potentially reduce the importance of lensing effects compared to other systematics. 

The formalism we have designed here is applicable to many different intensity-mapping surveys.  Specifically, it can be applied as is to any survey where all of the emitting sources are small compared to the voxel size.  For most target lines, all of the emission comes from within galaxies, and most intensity maps will have resolutions much larger than any galaxy, so this assumption holds.  However, this is not the case for measurements of the 21 cm line at the epoch of reionization, as this emission comes from large volumes of diffuse intergalactic gas.  Ly$\alpha$ surveys may have a similar IGM component as well \citep{Pullen2014}, though the intensity of this component is a subject of debate.  For diffuse emission, the concept of a luminosity function is not well defined, and our formalism breaks down.  The one-point statistics of such lines must be treated in a very different manner, such as the methods described in \citet{Barkana2008} or \citet{Shimabukuro2015}.

Our results here provide an excellent proof-of-concept of the VID method.  However, there are a number of subtle effects that require deeper study in future work.  One such effect is due to the clustering of the source galaxies.  We have treated the different bins of the voxel histogram as entirely independent, which may not be the case in a real, clustered map.  If a map contains a large over density, for example, we would expect to see an excess of voxels in several bright bins.  The opposite is true for a large underdensity.  Similarly, we have neglected any effects of beam smoothing.  If the beam is large compared to the chosen voxel size, then some emission from a galaxy in one voxel will be smeared into adjacent voxels, leading to correlations between nearby voxels.  We chose voxels of comparable size to the beam, so this effect should not be dramatic, but it deserves further analysis.  Many other instrumental systematics, such as ground contamination and pointing errors may also degrade our constraints.

One way to test these and other effects in greater detail would be to make use of maps simulated from large N-body codes, such as those used in \citet{Li2016}.  Many components of our VID calculation, such as the halo bias and the galaxy number count distribution could be easily studied based on such simulations.  It would also be substantially easier to test clustering, beam smoothing, and line-of-sight mode subtraction in a simulated map than in our analytic work.  We used the simulations from \citet{Breysse2015} in Appendix \ref{Sim} to test the numerical stability of our VID calculations, but these simulations do not include many of the effects described here.  Full-scale numerical simulations will be an important tool as we prepare to apply this formalism to real data.

\section{Conclusion}
We have presented a powerful new method for measuring line luminosity functions from intensity maps using the probability distribution of voxel intensities.  This voxel intensity distribution can be calculated using $P(D)$ analysis techniques and measured from a map by making a histogram of voxel intensities.  Because intensity maps are extremely non-Gaussian, this one-point statistic contains a substantial amount of information that cannot be obtained from usual power spectrum analyses.  

We tested our formalism on a four-parameter model of CO emission observed by an experiment similar to the planned COMAP survey.  We found that the VID statistic was able to constrain these four parameters with an average error of order $\sim10\%$, despite not including any prior information.  Incorporating various forms of foreground contamination such as continuum emission, interloper lines, and gravitational lensing weakens these constraints by varying degrees.  However,  the VID statistic still provides useful information despite these contaminants, even in very pessimistic cases where the power spectrum would be completely swamped by foregrounds.  Our results here serve as an excellent proof of the VID concept.  Though more work is necessary to fine tune the various subtleties of this method, this work suggests that the VID will make a powerful addition to the intensity mapping toolbox as more and more experiments come online in the coming years.

The authors would like to thank Tony Li, Garrett Keating, and the participants of the Opportunities and Challenges in Intensity Mapping Workshop for useful discussions.  This work was supported at JHU by NSF Grant No. 0244990, NASA NNX15AB18G, the John Templeton Foundation, and the Simons Foundation.  LD is supported at the Institute for Advanced Study by NASA through Einstein Postdoctoral Fellowship grant number PF5-160135 awarded by the Chandra X-ray Center, which is operated by the Smithsonian Astrophysical Observatory for NASA under contract NAS8-03060.  PSB was supported by program number HST-HF2-51353.001-A, provided by NASA through a Hubble Fellowship grant from the Space Telescope Science Institute, which is operated by the Association of Universities for Research in Astronomy, Incorporated, under NASA contract NAS5-26555.

\appendix

\section{Effects of luminosity-dependent bias}
\label{bias}
As mentioned above, though our formalism takes into account the average halo bias when computing $\Pn$, we do not correctly include the full luminosity dependence of the bias.  Here we will describe an approximate method for computing the full biased VID and attempt to get an idea of how important the effect is.  Since our fiducial model is not derived from any sort of mass-luminosity relation, for the purposes of this discussion we will look at the probability distribution $\p(M)$ of total halo mass contained within a voxel rather than total intensity.  We will assume that halo masses are drawn from the Tinker mass function \citep{Tinker2008} and we will use the corresponding $b(M)$ from \citet{Tinker2010}.

In our normal formalism, we would compute $\sigma_G$ using the average bias $\overline{b}$, which when calculating $\p(M)$ would be 
\be
\overline{b}=\frac{\int b(M) dn/dM dM}{\int{dn/dM dM}}.
\ee
Now, instead of computing one average bias for all galaxies, assume instead that we split our population in half around some mass value $M_{\rm{edge}}$.  We can then compute two average biases, two $\sigma_G$'s, and two mass PDFs $\p_i(M)$ for the low- and high-mass populations.  By the same logic used in Equation (\ref{P2}), the full $\p(M)$ will be the convolution of those of the two subsets.  We can then divide our mass range into smaller and smaller subsets to more accurately model a full continuous $b(M)$.  

\begin{figure}
\centering
\includegraphics[width=\columnwidth]{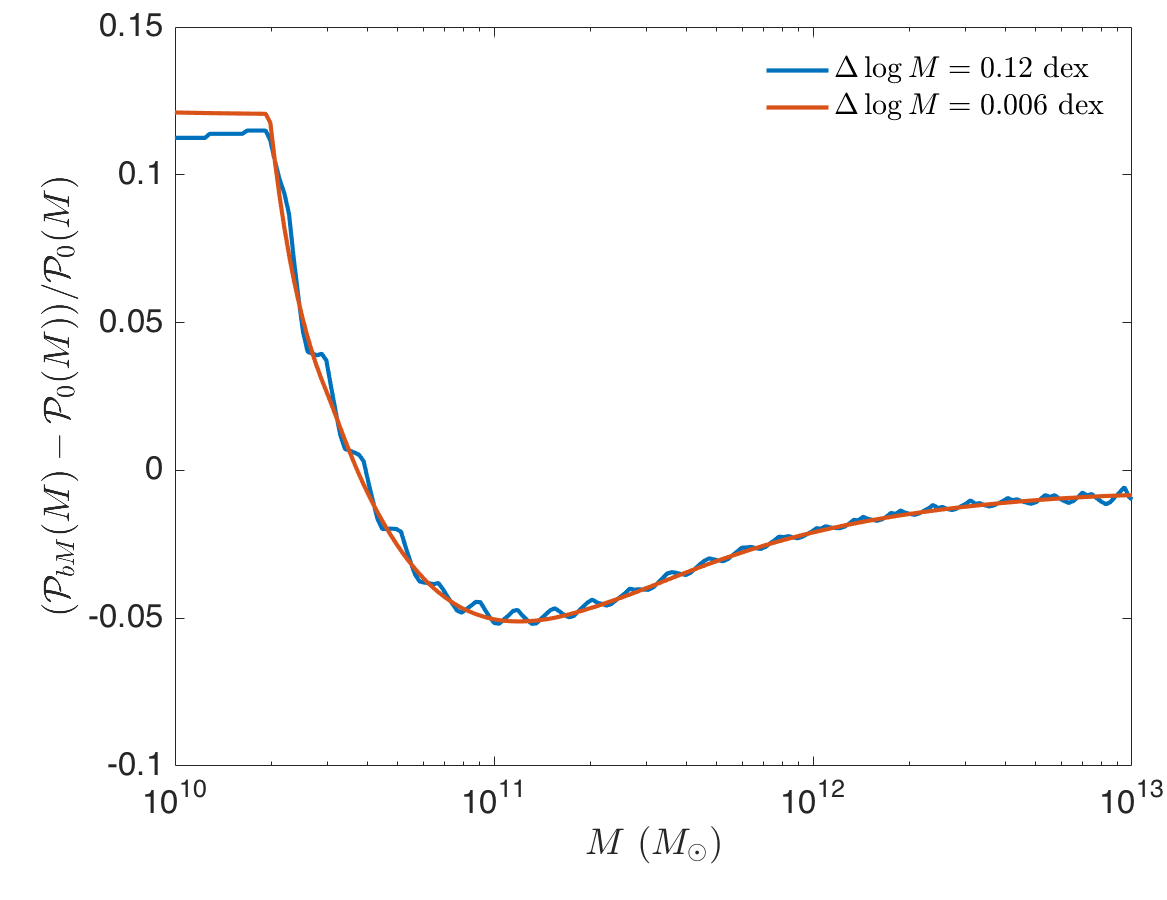}
\caption{Fractional change in $\p(M)$ when including mass-dependent bias in wider bins (blue) ad narrower bins(red).}
\label{bM}
\end{figure}

Figure \ref{bM} shows the fractional change in $\p(M)$ from a single average bias bin to many narrow bins.  If the bins are too wide, there are significant discontinuities at the bin edges, but as the bins become smaller we can see that these edge effects become small.  At the low mass end, the effect of this bias is of order $\sim10\%$, however these low-mass voxels would likely fall into the noise-dominated regime of a full VID calculation.  This effect is small enough ($\lesssim5\%$ outside the noise-dominated region) that we do not expect a full Fisher analysis including the effects of bias to produce significantly different constraints.  However, this does imply that leaving this effect out of the analysis of future measurements could lead to non-negligible systematic errors.  It is therefore important for future models to accurately take into account the luminosity-dependent bias.

\section{Simulated VIDs}
\label{Sim}
\citet{Breysse2015} demonstrated a method for simulating 2D slices of intensity maps with given power spectra.  Our VID formalism is based in part on these simulations, and the two methods are based on the same set of assumptions.  Both assume galaxies have randomly assigned luminosities drawn from a luminosity function and are distributed on the sky according to a lognormal random field.  If we estimate a VID from a simulation prepared using the \citet{Breysse2015} routine, we can then verify that the calculations presented here produce reliable results.

\begin{figure}
\centering
\includegraphics[width=\columnwidth]{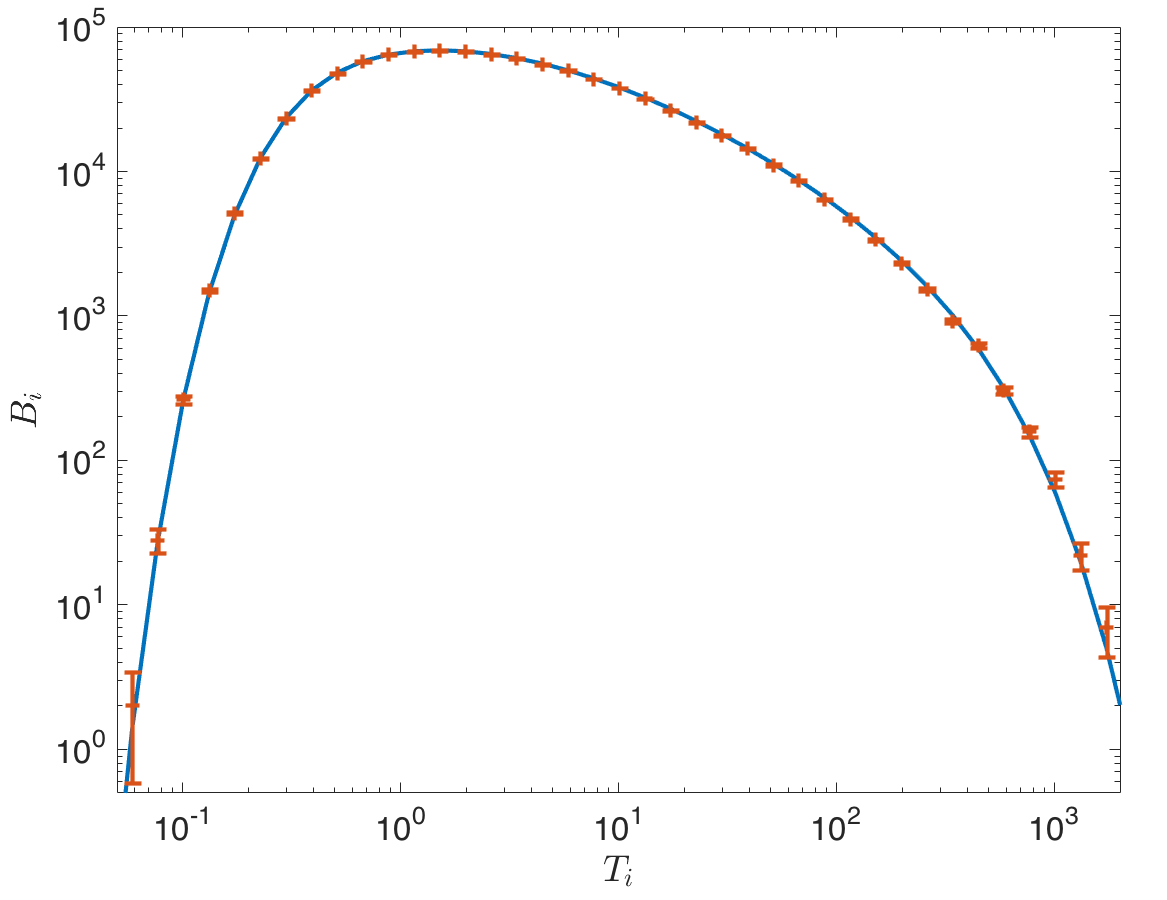}
\caption{Number of voxels in intensity bins $B_i$ simulated using the \citet{Breysse2015} method (red points) compared with the predicted values from our fiducial CO VID (blue).  Error bars on the simulated bins are Poisson.}
\label{SimCompare}
\end{figure}

We generate 400 slices of a CO intensity map with galaxy luminosities drawn from our fiducial $\Phi(L)$ with voxel sizes set by the COMAP parameters.  We then bin the resulting map in bins of width $\Delta \log T=0.12$ dex, and compare to the bin values predicted by our fiducial VID.  Figure \ref{SimCompare} shows the results.  The plotted error bars on the simulated bins are the Poisson errors we use when computing Fisher matrices.  This result clearly shows that our simulations and VID calculations are in good agreement, and that our numerical computations do not introduce a significant amount of error into our final VIDs.  The simulations still use the same set of approximations we used when deriving the VIDs though (see discussions above), so it would be useful in the future to test our formalism against more in-depth N-body simulations which would take into account, for example, the full mass-dependent halo bias.

\section{Full FG2 Constraints}
\begin{figure*}
\centering
\includegraphics[width=\textwidth]{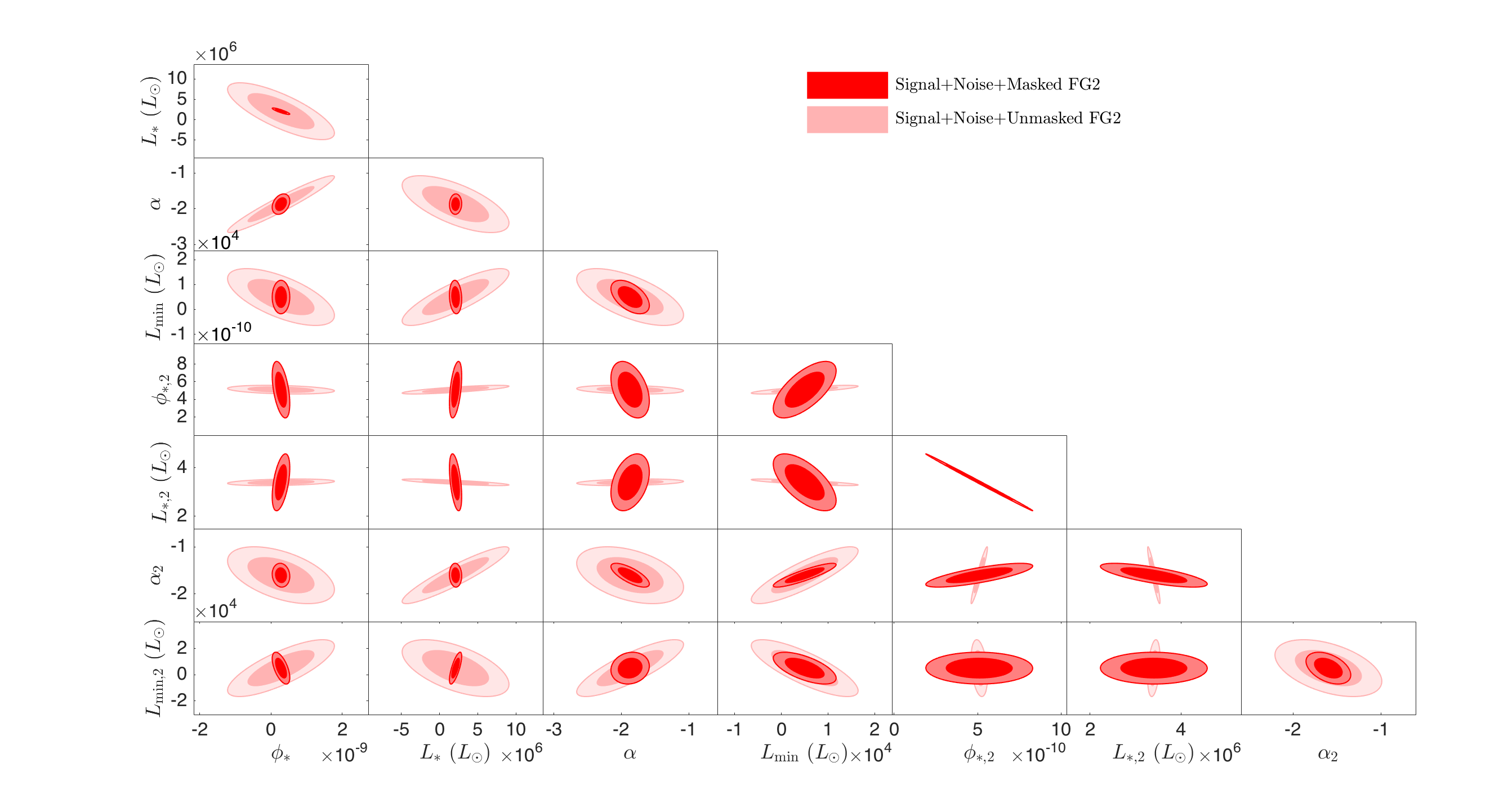}
\caption{Full Fisher matrix constraints for a model including the CO signal, the COMAP instrumental noise, and clustering foreground FG2 both masked (dark) and unmasked (light).  Parameters with subscript ``2" denote parameters of the FG2 luminosity function.}
\label{FG2_Full}
\end{figure*}

Figure \ref{FG2_Full} shows the full 8-parameter Fisher matrix constraints on the parameters of our fiducial CO model and those of the clustering foreground FG2.  Light red ellipses show the constraints with the unmasked foreground, dark ellipses show those with the masking from Figure \ref{FGs} applied.  The parameter constraints change in some counter-intuitive ways when the masking is applied, due to the substantial degeneracies between the signal and foreground parameters in the unmasked case.  For example, the constraints in the $[L_*,L_{*,2}]$ parameter space rotate by nearly 90 degrees when masked.  This occurs because the masking removes nearly all of the information about the foreground cutoff, while simultaneously leaving the signal cutoff free of contamination.

\section{Luminosity Function Constraints}
\label{LFerrApp}
\begin{figure}
\centering
\includegraphics[width=\columnwidth]{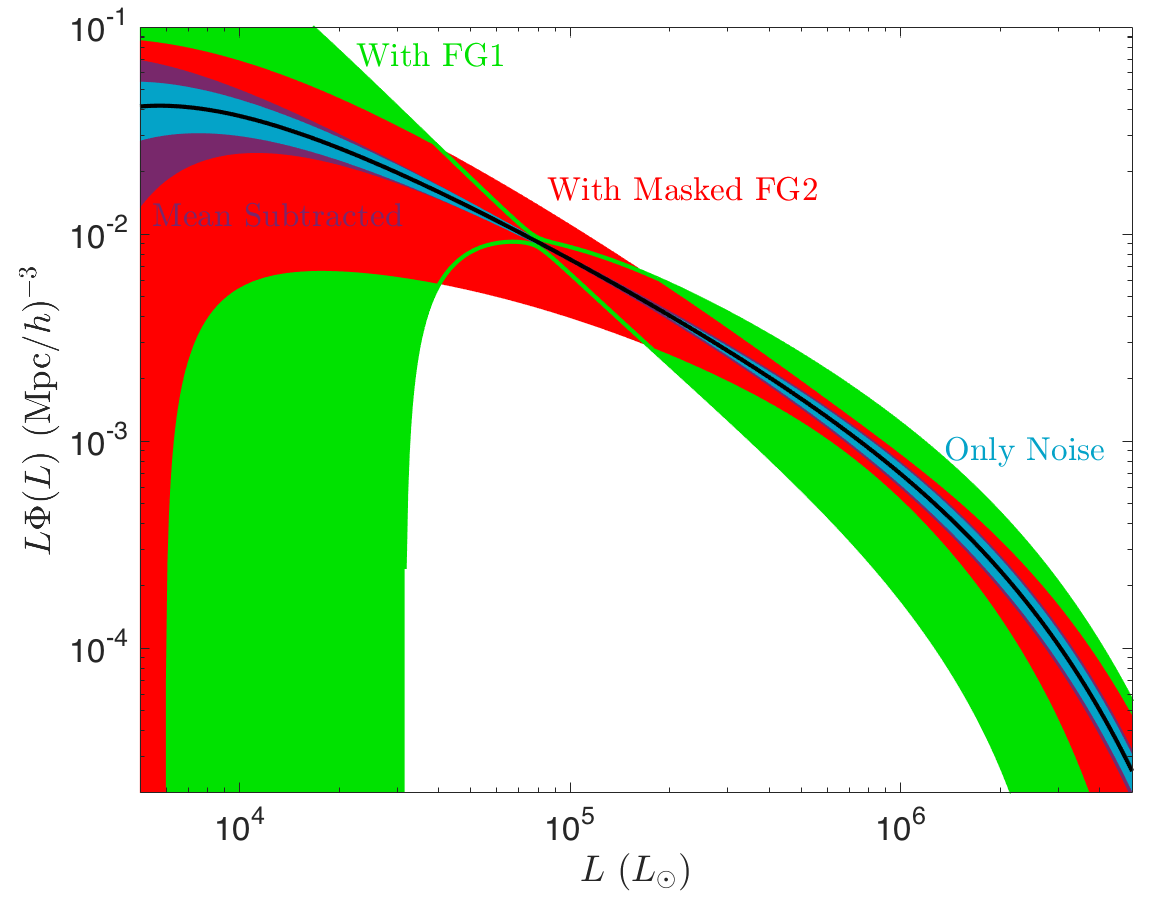}
\caption{95\% confidence regions around our fiducial luminosity function for cases with only instrumental noise (blue), noise plus mean subtraction (purple), noise plus shot-noise foreground FG1 (green), and noise plus masked clustering foreground FG2 (red).}
\label{LFerrBig}
\end{figure}

Just as we did in Figure \ref{LFerrFig}, we can use our Fisher matrix constraints on the fiducial luminosity function parameters under various forms of contamination to estimate uncertainties on the luminosity function $\Phi(L)$.  The results of this procedure are shown in Figure \ref{LFerrBig}.  The region shown in blue give the same constraints shown previously for the case including only signal and instrumental noise.  The purple region shows the effect of subtracting out the mean of the map to remove continuum foregrounds.  As described previously, the constraints worsen somewhat but on the whole do not change substantially.

The green region shows the constraints with the shot-noise FG1 line added.  The constraints get notably worse when this contamination is added, but the VID remains reasonably constrained above a few times $10^4$ $L_{\sun}$ despite the fact that no attempt has been made to clean out the interloper line.  The errors reach a minimum at around $L\sim10^5\ L_{\sun}$.  This is roughly the luminosity where the difference between the signal and FG1 luminosity functions is greatest.  It is also just above the point where the instrumental noise falls off, so the signal-to-noise is maximized.  Errors with a masked clustering foreground FG2 are shown in red.  However, once we mask out the brightest foreground sources, we see that we get significant constraining power over nearly the entire luminosity range, with the best constraints coming at high $L$ where the foreground is masked out.


\begin{thebibliography}{999}

\bibitem[\protect\citeauthoryear{Ali et al.}{2015}]{Ali2015} Ali Z.~S., et al., 2015, ApJ, 809, 61 (arXiv:1502.06016)

\bibitem[\protect\citeauthoryear{Bandura et al.}{2014}]{Bandura2014} Bandura K., et al., 2014, SPIE, 9145, 914522 (arXiv:1406.2288) 

\bibitem[\protect\citeauthoryear{Barcons}{1992}]{Barcons1992} 
Barcons X., 1992, ApJ, 396, 460 

\bibitem[\protect\citeauthoryear{Barcons et al.}{1994}]{Barcons1994} Barcons X., Raymont G.~B., Warwick R.~S., Fabian A.~C., Mason K.~O., McHardy I., Rowan-Robinson M., 1994, MNRAS, 268, 833 

\bibitem[\protect\citeauthoryear{Barkana \& Loeb}{2008}]{Barkana2008} Barkana R., Loeb A., 2008, MNRAS, 384, 1069 (arXiv:0705.3246) 

\bibitem[\protect\citeauthoryear{Bartolo, Matarrese, \& Riotto}{2010}]{Bartolo2010} Bartolo N., Matarrese S., Riotto A., 2010, AdAst, 2010, 157079 (arXiv:1001.3957) 

\bibitem[Behroozi et al.(2013)]{Behroozi2013} Behroozi, P.~S., Wechsler, R.~H., \& Conroy, C.\ 2013, \apj, 770, 57 (arXiv:1207.6105) 

\bibitem[Breysse et al.(2014)]{Breysse2014} Breysse, P.~C., Kovetz, E.~D., \& Kamionkowski, M.\ 2014, \mnras, 443, 3506 (arXiv:1405.0489) 

\bibitem[Breysse et al.(2015)]{Breysse2015} Breysse, P.~C., Kovetz, E.~D., \& Kamionkowski, M.\ 2015, \mnras, 452, 3408 (arXiv:1503.05202) 

\bibitem[Breysse et al.(2016)]{Breysse2016a} Breysse, P.~C., Kovetz, E.~D., \& Kamionkowski, M.\ 2016, \mnras, 457, L127 (arXiv:1507.06304) 

\bibitem[\protect\citeauthoryear{Breysse \& Rahman}{2016}]{Breysse2016b} Breysse P.~C., Rahman M., 2016, arXiv:1606.07820 

\bibitem[\protect\citeauthoryear{Cheng et al.}{2016}]{Cheng2016} Cheng Y.-T., Chang T.-C., Bock J., Bradford C.~M., Cooray A., 2016, arXiv, arXiv:1604.07833 

\bibitem[\protect\citeauthoryear{Coles \& Jones}{1991}]{Coles1991} Coles P., Jones B., 1991, MNRAS, 248, 1

\bibitem[\protect\citeauthoryear{Comaschi, Yue, \& Ferrara}{2016}]{Comaschi2016} Comaschi P., Yue B., Ferrara A., 2016, arXiv:1605.05733 

\bibitem[\protect\citeauthoryear{Carilli \& Walter}{2013}]{Carilli2013} Carilli C.~L., Walter F., 2013, ARA\&A, 51, 105 (arXiv:1301.0371) 

\bibitem[\protect\citeauthoryear{Crites et al.}{2014}]{Crites2014} Crites A.~T., et al., 2014, SPIE, 9153, 91531W 

\bibitem[\protect\citeauthoryear{Das \& Ostriker}{2006}]{Das2006} Das S., Ostriker J.~P., 2006, ApJ, 645, 1 (arXiv:0512644) 

\bibitem[\protect\citeauthoryear{DeBoer et al.}{2016}]{DeBoer2016} DeBoer D.~R., et al., 2016, arXiv:1606.07473 

\bibitem[\protect\citeauthoryear{Dor{\'e} et al.}{2014}]{Dore2014} Dor{\'e} O., et al., 2014, arXiv:1412.4872 

\bibitem[\protect\citeauthoryear{Fonseca et al.}{2016}]{Fonseca2016} Fonseca J., Silva M., Santos M.~G., Cooray A., 2016, arXiv:1607.05288 

\bibitem[\protect\citeauthoryear{Glenn et al.}{2010}]{Glenn2010} Glenn J., et al., 2010, MNRAS, 409, 109 (arXiv:1009.5675) 

\bibitem[\protect\citeauthoryear{Gong et al.}{2011}]{Gong2011} Gong Y., Chen X., Silva M., Cooray A., Santos M.~G., 2011, ApJ, 740, L20 (arXiv:1108.0947) 

\bibitem[\protect\citeauthoryear{Gong et al.}{2012}]{Gong2012} Gong Y., Cooray A., Silva M., Santos M.~G., Bock J., Bradford C.~M., Zemcov M., 2012, ApJ, 745, 49 (arXiv:1107.3553)

\bibitem[\protect\citeauthoryear{Gong, Cooray, \& Santos}{2013}]{Gong2013} Gong Y., Cooray A., Santos M.~G., 2013, ApJ, 768, 130 (arXiv:1212.2964) 

\bibitem[\protect\citeauthoryear{Gong et al.}{2014}]{Gong2014} Gong Y., Silva M., Cooray A., Santos M.~G., 2014, ApJ, 785, 72 (arXiv:1312.2035) 

\bibitem[\protect\citeauthoryear{Kennicutt}{1998}]{Kennicutt1998} Kennicutt R.~C., Jr., 1998, ApJ, 498, 541 (arXiv:9712213) 

\bibitem[\protect\citeauthoryear{Jungman et al.}{1996a}]{Jungman1996a} Jungman G., Kamionkowski M., Kosowsky A., Spergel D.~N., 1996, PhRvL, 76, 1007 (arXiv:9507080)

\bibitem[\protect\citeauthoryear{Jungman et al.}{1996b}]{Jungman1996b} Jungman G., Kamionkowski M., Kosowsky A., Spergel D.~N., 1996, PhRvD, 54, 1332 (arXiv:9512139) 

\bibitem[\protect\citeauthoryear{Kayo, Taruya, \& Suto}{2001}]{Kayo2001} Kayo I., Taruya A., Suto Y., 2001, ApJ, 561, 22 (arXiv:0105218) 

\bibitem[\protect\citeauthoryear{Keating et al.}{2015}]{Keating2015} Keating G.~K., et al., 2015, ApJ, 814, 140 (arXiv:1510.06744) 

\bibitem[\protect\citeauthoryear{Keating et al.}{2016}]{Keating2016} Keating G.~K., Marrone D.~P., Bower G.~C., Leitch E., Carlstrom J.~E., DeBoer D.~R., 2016, arXiv:1605.03971 

\bibitem[\protect\citeauthoryear{Lee et al.}{2009}]{Lee2009} Lee S.~K., Ando S., Kamionkowski M., 2009, JCAP, 7, 007 (arXiv:0810.1284) 

\bibitem[\protect\citeauthoryear{Lee et al.}{2015}]{Lee2015} Lee S.~K., Lisanti M., Safdi B.~R., 2015, JCAP, 5, 056 (arXiv:1412.6099) 

\bibitem[\protect\citeauthoryear{Lewis \& Challinor}{2011}]{Lewis2011} Lewis A., Challinor A., 2011, Astrophysics Source Code Library, record ascl:1102.026

\bibitem[Li et al.(2016)]{Li2016} Li, T.~Y., Wechsler, R.~H., Devaraj, K., \& Church, S.~E.\ 2016, \apj, 817, 169 (arXiv:1503.08833)

\bibitem[\protect\citeauthoryear{Lidz et al.}{2011}]{Lidz2011} Lidz A., Furlanetto S.~R., Oh S.~P., Aguirre J., Chang T.-C., Dor{\'e} O., Pritchard J.~R., 2011, ApJ, 741, 70 (arXiv:1104.4800) 

\bibitem[\protect\citeauthoryear{Lidz \& Taylor}{2016}]{Lidz2016} Lidz A., Taylor J., 2016, arXiv:1604.05737 (arXiv:1604.05737)

\bibitem[\protect\citeauthoryear{Limber}{1953}]{Limber1953} Limber D.~N., 1953, ApJ, 117, 134 

\bibitem[\protect\citeauthoryear{Morales \& Wyithe}{2010}]{Morales2010} Morales M.~F., Wyithe J.~S.~B., 2010, ARA\&A, 48, 127 (arXiv:0910.3010) 

\bibitem[\protect\citeauthoryear{Newburgh et al.}{2016}]{Newburgh2016} Newburgh L.~B., et al., 2016, arXiv:1607.02059 

\bibitem[\protect\citeauthoryear{Planck Collaboration et al.}{2015a}]{Planck2015a} Planck Collaboration, et al., 2015, arXiv:1502.01588 

\bibitem[\protect\citeauthoryear{Planck Collaboration et al.}{2015b}]{Planck2015} Planck Collaboration, et al., 2015, arXiv:1502.01592 

\bibitem[\protect\citeauthoryear{Planck Collaboration et al.}{2015c}]{Planck2015c} Planck Collaboration, et al., 2015, arXiv:1507.02704 

\bibitem[Popping et al.(2016)]{Popping2016} Popping, G., van Kampen, E., Decarli, R., et al.\ 2016, arXiv:1602.02761

\bibitem[\protect\citeauthoryear{Pullen et al.}{2013}]{Pullen2013}  Pullen A.~R., Chang T.-C., Dor{\'e} O., Lidz A., 2013, ApJ, 768, 15 (arXiv:1211.1397) 

\bibitem[\protect\citeauthoryear{Pullen, Dor{\'e}, \& Bock}{2014}]{Pullen2014} Pullen A.~R., Dor{\'e} O., Bock J., 2014, ApJ, 786, 111 (arXiv:1309.2295) 

\bibitem[\protect\citeauthoryear{Righi, Hern{\'a}ndez-Monteagudo, \& Sunyaev}{2008}]{Righi2008} Righi M., Hern{\'a}ndez-Monteagudo C., Sunyaev R.~A., 2008, A\&A, 489, 489 (arXiv:0805.2174)

\bibitem[\protect\citeauthoryear{Rubin}{1954}]{Rubin1954} Rubin V.~C., 1954, PNAS, 40, 541

\bibitem[\protect\citeauthoryear{Santos et al.}{2015}]{Santos2015} Santos M., et al., 2015, Proc. of Advancing Astrophysics with the SKA. Giardini Naxos, Italy. id.19 (arXiv:1501.03989)

\bibitem[\protect\citeauthoryear{Schechter}{1976}]{Schechter1976} 
Schechter P., 1976, ApJ, 203, 297

\bibitem[\protect\citeauthoryear{Scheuer}{1957}]{Scheuer1957} Scheuer P.~A.~G., 1957, PCPS, 53, 764 

\bibitem[\protect\citeauthoryear{Shimabukuro et al.}{2015}]{Shimabukuro2015} Shimabukuro H., Yoshiura S., Takahashi K., Yokoyama S., Ichiki K., 2015, MNRAS, 451, 467 (arXiv:1412.3332) 

\bibitem[\protect\citeauthoryear{Silva et al.}{2015}]{Silva2015} Silva M., Santos M.~G., Cooray A., Gong Y., 2015, ApJ, 806, 209 (arXiv:1410.4808) 

\bibitem[\protect\citeauthoryear{Smith et al.}{2003}]{Smith2003} Smith R.~E., et al., 2003, MNRAS, 341, 1311 (arXiv:0207664)

\bibitem[\protect\citeauthoryear{Suginohara, Suginohara, \& Spergel}{1999}]{Suginohara1999} Suginohara M., Suginohara T., Spergel D.~N., 1999, ApJ, 512, 547 (arXiv:astro-ph/9803236)

\bibitem[\protect\citeauthoryear{Takahashi et al.}{2011}]{Takahashi2011} Takahashi R., Oguri M., Sato M., Hamana T., 2011, ApJ, 742, 15 (arXiv:1106.3823)

\bibitem[\protect\citeauthoryear{Tegmark et al.}{1998}]{Tegmark1998} Tegmark M., Hamilton A.~J.~S., Strauss M.~A., Vogeley M.~S., Szalay A.~S., 1998, ApJ, 499, 555 (arXiv:9708020)

\bibitem[\protect\citeauthoryear{Tingay et al.}{2013}]{Tingay2013} Tingay S.~J., et al., 2013, PASA, 30, e007 (arXiv:1206.6945) 

\bibitem[\protect\citeauthoryear{Tinker et al.}{2008}]{Tinker2008} Tinker J., Kravtsov A.~V., Klypin A., Abazajian K., Warren M., Yepes G., Gottl{\"o}ber S., Holz D.~E., 2008, ApJ, 688, 709 (arXiv:0803.2706) 

\bibitem[Tinker et al.(2010)]{Tinker2010} Tinker, J.~L., Robertson, B.~E., Kravtsov, A.~V., et al.\ 2010, \apj, 724, 878 (arXiv:1001.3162)

\bibitem[\protect\citeauthoryear{van Haarlem et al.}{2013}]{Haarlem2013} van Haarlem M.~P., et al., 2013, A\&A, 556, A2 (arXiv:1305.3550) 

\bibitem[\protect\citeauthoryear{Visbal \& Loeb}{2010}]{Visbal2010} Visbal E., Loeb A., 2010, JCAP, 11, 16 (arXiv:1008.3178)

\bibitem[\protect\citeauthoryear{Visbal, Haiman, \& Bryan}{2015}]{Visbal2015} Visbal E., Haiman Z., Bryan G.~L., 2015, MNRAS, 450, 2506 (arXiv:1501.03177)

\bibitem[\protect\citeauthoryear{Visbal, Trac, \& Loeb}{2011}]{Visbal2011} Visbal E., Trac H., Loeb A., 2011, JCAP, 8, 10 (arXiv:1104.4809)

\bibitem[\protect\citeauthoryear{Yue et al.}{2015}]{Yue2015} Yue B., Ferrara A., Pallottini A., Gallerani S., Vallini L., 2015, MNRAS, 450, 3829 (arXiv:1504.06530)

\end{thebibliography}
\end{document}